\documentclass[sigconf]{acmart}

\copyrightyear{2020}
\acmYear{2020}
\setcopyright{acmlicensed}
\acmConference[CCS '20] {2020 ACM SIGSAC Conference on Computer and Communications Security}{November 9--13, 2020}{Virtual Event, USA}
\acmBooktitle{2020 ACM SIGSAC Conference on Computer and Communications Security (CCS '20), November 9--13, 2020, Virtual Event, USA}
\acmPrice{15.00}
\acmDOI{10.1145/3372297.3423345}
\acmISBN{978-1-4503-7089-9/20/11}

\usepackage{booktabs}
\usepackage[T1]{fontenc}
\usepackage{graphicx}
\usepackage{hyperref}
\hypersetup{
  colorlinks = true,
  linkcolor = blue,
  anchorcolor = blue,
  citecolor = blue,
  filecolor = blue,
  urlcolor = blue
}
\usepackage[utf8]{inputenc}
\usepackage{xspace}           %

\usepackage{fontawesome}
\usepackage{tikz}
\usepackage[group-digits=true, group-minimum-digits = 4, group-separator={,}, scientific-notation=engineering]{siunitx}
\newif\iflongversion\longversiontrue

\iflongversion
    \usepackage{hyperref}
    \usepackage{tikz}
    \newcommand\copyrighttext{
        \vspace*{3pt}
        \hfill Extended Paper Version \hfill\hfill\\[4pt]
        \footnotesize
        If you cite this paper, please use the CCS reference:
        Jens Hiller, Johanna Amann, and Oliver Hohlfeld. 2020. The Boon and Bane of Cross-Signing: Shedding Light on a Common Practice in Public Key Infrastructures. In 2020 ACM SIGSAC Conference on Computer and Communications Security (CCS ’20), November 9–13, 2020, Virtual Event, USA. ACM, New York, NY, USA, 18 pages. https://doi.org/10.1145/3372297.3423345
        \vspace*{1pt}
    }
    \newcommand\copyrightnotice{%
    \begin{tikzpicture}[remember picture,overlay]
    \node[anchor=south,yshift=-65pt,draw] at (current page.north) {\parbox{\dimexpr\textwidth-\fboxsep-\fboxrule\relax}{\copyrighttext}};
    \end{tikzpicture}
    }
\fi

\begin{document}

\iflongversion
\else
  \fancyhead{}  %
\fi

\title{The Boon and Bane of Cross-Signing: Shedding Light on a Common Practice in Public Key Infrastructures}

\author{Jens Hiller}
\authornote{Parts of the work conducted during an internship at the International Computer Science Institute (ICSI).}
\email{hiller@comsys.rwth-aachen.de}
\affiliation{%
  \institution{Communication and Distributed Systems, RWTH Aachen University}
}
\author{Johanna Amann}
\email{johanna@icir.org}
\affiliation{%
  \institution{ICSI; Corelight; LBNL}
}
\author{Oliver Hohlfeld}
\email{hohlfeld@b-tu.de}
\affiliation{%
  \institution{Brandenburg University of Technology}
}

\newcommand{\etal}{et~al.\xspace}

\newcommand{\xsc}[1]{XS-Cert}
\newcommand{\xscs}[1]{\xsc{}s}

\newcommand{\todo}[1]{\textbf{\color{red}{TODO: #1}}}
\newcommand{\fix}[1]{{\color{red}{#1}}}

\newcommand{\ca}[2]{#1}

\newif\iffontawesomesymbols\fontawesomesymbolstrue

\iffontawesomesymbols
\newcommand{\symbolBadXS}{{\normalsize\faWarning}}
\newcommand{\symbolGoodXS}{{\normalsize\faCheckCircleO}}
\newcommand{\symbolNeutralXS}{\tikz[baseline=-1.5pt]{\node[inner sep=0pt] at (0pt,+2pt) {\bf +}; \node[inner sep=0pt] at (3.5pt,+0.8pt) {\bf $\diagup$}; \node[inner sep=0pt] at (6pt,-1.5pt) {\bf -};}}  %
\newcommand{\symbolUglyXS}{{\normalsize\faBug}}

\else

\newcommand{\symbolBadXS}{{$\bf\mathcal{B}$}}  %
\newcommand{\symbolGoodXS}{{\bf\normalsize\checkmark}}  %
\newcommand{\symbolNeutralXS}{\tikz[baseline=-1.5pt]{\node[inner sep=0pt] at (0pt,+2pt) {\bf +}; \node[inner sep=0pt] at (3.5pt,+0.8pt) {\bf $\diagup$}; \node[inner sep=0pt] at (6pt,-1.5pt) {\bf -};}}  %
\newcommand{\symbolUglyXS}{{\bf$\mathcal{U}$}}  %
\fi

\begin{abstract}
Public Key Infrastructures (PKIs) with their trusted Certificate Authorities (CAs) provide the trust backbone for the Internet:
CAs sign certificates which prove the identity of servers, applications, or users.
To be trusted by operating systems and browsers, a CA has to undergo lengthy and costly validation processes.
Alternatively, trusted CAs can \emph{cross-sign} other CAs to extend their trust to them.
In this paper, we systematically analyze the present and past state of cross-signing in the Web PKI.
Our dataset (derived from passive TLS monitors \emph{and} public CT logs) encompasses more than 7 years and 225 million certificates with 9.3 billion trust paths.
We show benefits and risks of cross-signing.
We discuss the difficulty of revoking trusted CA certificates where, worrisome, cross-signing can result in valid trust paths to remain \emph{after} revocation; a problem for non-browser software that often blindly trusts all CA certificates and ignores revocations.
However, cross-signing also enables fast bootstrapping of new CAs, e.g., Let's Encrypt, and achieves a non-disruptive user experience by providing backward compatibility. %
In this paper, we propose new rules and guidance for cross-signing to preserve its positive potential while mitigating its risks.

\end{abstract}


\maketitle

\iflongversion
  \copyrightnotice
\fi

\section{Introduction} %
\label{sec:introduction}

Public key infrastructures (PKIs) like the Web PKI, provide the trust infrastructure for many applications in today's Internet.
They, e.g., enable webbrowsers, or apps on mobile operating systems (OS), to authenticate servers for secure online banking, web shopping, or password entry.
Governments use PKIs for authentication in privacy-preserving health systems, remote functionality of administrative offices, or electronic voting~\cite{5374363,7001845,5509037,6399701}.

Certificate Authorities (CAs) serve as trust anchors in PKIs and have the ability to issue trusted certificates to companies and individuals.
The security of a PKI relies on benign and correct acting of \emph{all} its CAs.
Despite audit processes, there have been several cases of severe CA misbehavior or security breaches:
In 2011, the DigiNotar CA was compromised~\cite{prins2012_diginotar}. This caused its removal from root stores.
All DigiNotar issued certificates became untrusted.
In the following years, a range of new security measures were introduced to reduce the impact of future compromises~\cite{amann2017_mission}.
However, most face small deployment and thus have limited effect~\cite{amann2017_mission}.
Along this path, the \emph{Certification Authority Browser} (CAB) \emph{Forum} gradually increased the requirements that CAs must fulfill to remain in root stores.

Alternatively, trusted CAs can \emph{cross-sign} other CAs to extend their trust to them---thereby mitigating the lengthy and costly validation process that new CAs need to undergo.
Cross-signing describes the approach to obtain signatures from several issuers for one certificate\footnote{Technically, cross-signing creates several certificates that share subject and public key as each certificate has exactly one issuer.}.
It enables new CAs to quickly establish trust.
A prominent example is the bootstrapping of Let's Encrypt, which issued trusted certificates based on a cross-sign of their CA certificates by the already trusted CA IdenTrust while applying for root store inclusion of their own root certificate~\cite{letsencrypt-crosssign}.
Cross-signing also ensures broad validation of certificates in face of divergent root stores of OSes or applications.

However, cross-signing also bears risks: as cross-signs are not systematically tracked~\cite{roosa2013_trust}, cross-signing can challenge proper revocation of certificates in case of CA misbehavior, erroneous operation, or stolen keys.
The complexity added by cross-signs already resulted in too broad application of certificate revocation~\cite{globalsign_revocation_caused_issues_theRegister,globalsign_revocation_caused_issues_comodo}.
In this paper, we show that cross-signs also can lead to certificates remaining valid when their CA was distrusted; that the complexity of existing cross-signs makes revocation difficult; that different software and operating systems do not always thoroughly revoke certificates; and that cross-signing makes it difficult to track revocation of CA certificates, especially for non-browser software.

In this paper, we perform the first systematic study of the use and security effects of \emph{cross-signing} (also known as \emph{cross certification}), which is one major reason for missing transparency in PKIs~\cite{roosa2013_trust}.
For this, we use a passive TLS data-set that contains information about more than seven years of real-world TLS usage, containing more than \num{225} million certificates derived from more than \num{300} billion connections---which provides us with insights on the effect of cross-signs on real user connections.
For a broad coverage of CA certificates, we combine this private, user-centered dataset with publicly available data from Certificate Transparency (CT) logs.

The main contributions of our paper are as follows:

\begin{itemize}
  \item We provide a classification of different cross-sign patterns and analyze them with respect to their benefits, but also their risks of unexpected effects on the trust system.
  \item We systematically analyze the use of cross-signing in PKI systems, with a focus on the Web PKI. %
  Thereby, we reveal problematic cross-signs that render certificate revocation or root store removals ineffective, leading to unwanted valid trust paths.
  We also find legit use cases, e.g., cross-signing enabled the quick tremendous success of Let's Encrypt, eases the transition to progressive cryptography while maintaining compatibility for legacy applications, and makes a \emph{single} certificate trusted across different applications and operating systems, achieving a non-disruptive user experience.
  \item %
  We propose new rules and guidance for cross-signing to preserve its positive potential but mitigate enclosed risks.
\end{itemize}

\section{Background} %
\label{sec:certificate_basics}

This section gives a brief overview of how CAs establish trust, how trust is anchored in root stores, and how certificate revocation is applied today.
For a thorough description of PKIs and their fundamental concepts, we refer readers to \cite{holz2011_sslLandscape,clark2013_sok}.

\begin{figure}[tb]
  \centering
    \includegraphics[width=0.9\linewidth]{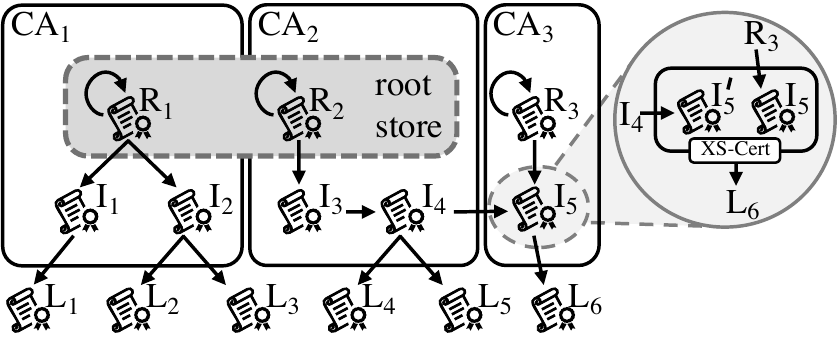}
  \caption{An example PKI including a cross-sign ($I_5 + I'_5$).}
  \label{fig:crosssign_certificates:definition}
  \Description[The figure shows a PKI including a cross-sign]{The figure shows root certificates which are included in a root store. These root certificates issue intermediate certificates which, in turn, issue leaf certificates. Furthermore, it shows that intermediate I5 was signed by the root R3 of CA3 as well as by the intermediate I4 of CA2. These two signatures technically lead to two certificates: I5 which is signed by R3 and I5' which is signed by I4. I5 and I5' form a XS-cert. This XS-cert, i.e., any of its certificates, provide valid signatures for the leaf certificate L6.}
\end{figure}

Operating systems and some web browsers maintain root stores. They serve as trust anchors when validating certificates:
To be valid, a certificate must be issued---directly or indirectly---by a trusted \emph{root certificate}, i.e., a certificate that is included in the root store.
Root certificates mostly issue \emph{intermediate certificates} which can issue further intermediate or \emph{leaf certificates}. %
Figure~\ref{fig:crosssign_certificates:definition} shows an example with several root ($R_i$), intermediate ($I_i$), and leaf ($L_i$) certificates.

\label{sec:certificate_basics:crl_methods}
In case of breaches like stolen private keys, certificates have to be revoked.
Revocation information is traditionally distributed by CAs using Certificate Revocation Lists (CRLs)~\cite{RFC3280} or interactively using OCSP~\cite{RFC2560}.
However, these mechanisms mostly remain unused due to (i) the overhead for distribution and (ii) inherent privacy concerns~\cite{chrome_disable_ocsp,prefetching}.
Applications thus often solely rely on the current state of the operating system's root store.

Consequently, Browsers and operating systems have started shipping vendor-controlled CRLs.
Mozilla uses \emph{OneCRL}~\cite{mozilla_onecrl_announcement}, Google \emph{CRLSets} \cite{chrome_crlsets} and a blacklist \cite{chrome_blocklist}; Microsoft and Apple include information on blocked CA certificates in their root stores \cite{crl_microsoft,crl_apple}.

\section{Cross-signing} %
\label{sec:crosssign_certificates}

We next introduce cross-signing and provide a classification of the different cross-signing patterns used in the remainder of the paper.

\subsection{Cross-signing: Definition}
\label{sec:crosssign_certificates:definition}

We illustrate cross-signing with a typical use-case:
To issue trusted certificates, a CA must be included in the respective root stores of web browsers, operating systems, etc.
Inclusion in these root stores requires time demanding audit and certification processes. %
During the process of being included in root stores, a CA may already want to issue trusted certificates.
To this end, another $CA_{\text{trusted}}$, whose certificate is already included in root stores, \emph{cross-signs} the CA's root or intermediate certificate to create a trust path that ends in the already trusted root certificate of $CA_{\text{trusted}}$.
In Figure~\ref{fig:crosssign_certificates:definition}, the intermediate certificate $I_5$ is cross-signed by $I_4$, providing a trust path to $R_2$.
As real-world example, Let's Encrypt has been using an intermediate certificate cross-signed by IdenTrust to already issue certificates while waiting for root store inclusion of its own root certificate~\cite{letsencrypt-crosssign}.
Similarly, CAs that are included in only some root stores can use cross-signing to extend trust to further root stores.
CAs typically call this \emph{cross-signing} (also \emph{cross-certification}) and the resulting certificates \emph{cross-certificates}~\cite{letsencrypt-crosssign,letsencrypt-crosssign-transition}.
Analogously, RFC~5280 defines a cross-certificate as a CA certificate that has different \emph{entities} as issuer and subject \cite{rfc5280}.
In this paper, we use a broader definition: (i) To analyze cross-signing for \emph{all} certificate types, i.e., root, intermediate, and leaf certificates, we consider all certificates, not only CA certificates.
(ii) To also track effects of signing a certificate with multiple CA certificates of the \emph{same} entity, we only require signatures by two different CA certificates, but not that issuer and subject are controlled by \emph{different} entities.

Specifically, our definition is as follows (cf. Figure~\ref{fig:crosssign_certificates:definition}).
To \emph{cross-sign} a certificate (here: $I_5$) that was originally issued by $R_3$,
a CA certificate (here: $I_4$) creates and signs a copy $I'_5$ which has the same subject and public key as $I_5$.
This process is necessary as each certificate has exactly one issuer field \cite{rfc5280}, i.e., $issuer(I_5) = R_3 ~\neq~ issuer(I'_5) = I_4$.
Thus, a cross-sign is a certificate for which another certificate exists that has the same subject and public key, but a different issuer and signature.
These certificates form a \emph{cross-sign certificate} (\xsc{}).
The certificates of a \xsc{} can be used interchangeably:
If $I_5$ and $I'_5$ are CA certificates, a certificate issued by $I_5$ will also validate using $I'_5$.
In more detail: when a certificate is validated, the validating software searches a CA certificate whose subject equals the issuer of the current certificate.
It then checks that the signature of the current certificate validates against the public key of the CA certificate.
As all certificates of a \xsc{} share subject and public key, $I_5$ and $I'_5$ can be used interchangeably.

Note that there is a difference between cross-signing and certificate \emph{re-issuances without re-keying}.
When a certificate reaches the end of its validity, it often is replaced by a certificate with the same subject, key and a new validity period.
We need to distinguish these cases from cross-signing.
Telling them apart is complicated since cross-signs often (and legitimately) do not have the exact same validity periods as the original certificate, e.g., because the \emph{not before} field should correspond to the issuance date \cite{mozilla_problematic_practices}.
\label{special:xs_validity_overlap_min}
As, e.g., GoDaddy allows for a renewal of certificates up to 120 days prior to expiration, we require certificates to have an overlapping validity period of at least \num{121} days to be considered cross-signs.

\begin{figure}[tb]
  \centering
    \includegraphics[]{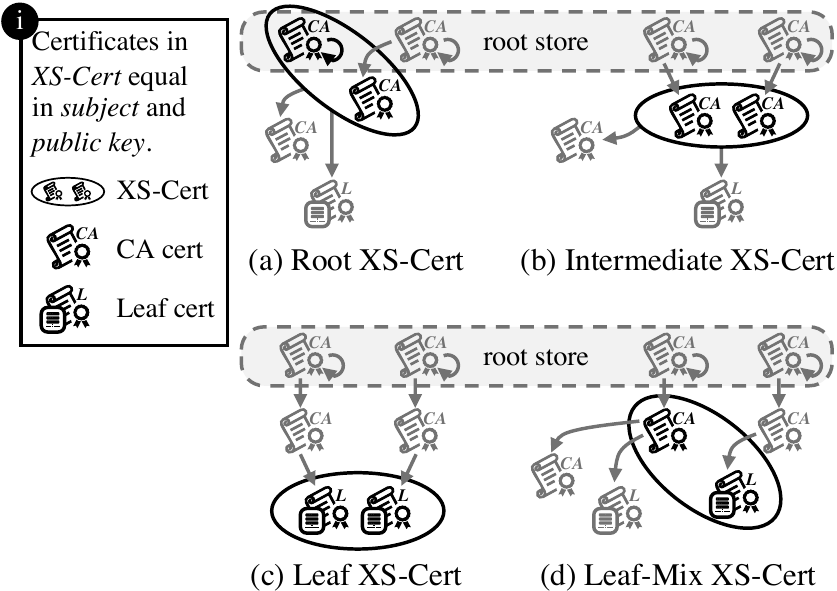}
  \caption{Cross-signing types. We observe many root, intermediate and leaf \xscs{} in our dataset. Luckily, leaf-mix \xscs{} remain a theory: We find no such problematic case.}
  \label{fig:crosssign_certificates:xs_types}
  \Description[The figure shows different cross-signing types.]{The figure conceptually shows root, intermediate, leaf and leaf-mix XS-certs as detailed in the main body of the paper.}
\end{figure}

\subsection{Cross-signing: Classification}
\label{sec:crosssign_certificates:classification}

For our analysis of cross-signing and corresponding risks for the PKI, we classify \xscs{} into 4 types. These types depend on the type of certificates in the \xsc{}. We illustrate this in Figure~\ref{fig:crosssign_certificates:xs_types}.

\textit{Root \xscs{}.}
\label{sec:crosssign_certificates:root-xs}
A root \xsc{} comprises at least two CA certificates of different issuers. At least one of them is part of a root store (cf. Figure~\ref{fig:crosssign_certificates:xs_types}a).
When CAs use the term cross-signing, they typically refer to this type.
This approach extends trust to root stores that do not include a CA's root certificate and can extend a CA's trust to more applications or operating systems.
Likewise, an already trusted root can bootstrap trust in a new CA certificate by cross-signing it.
Moreover, CAs can start to use a new CA certificate without disrupting compatibility with old applications that do not receive root store updates.
However, as we will discuss later, such cross-signs increase the complexity of removing trust in certificates that are part of a root store.
Trust paths can inadvertently remain valid via paths using other certificates of the \xsc{} (Section~\ref{sec:crossigns_problems:ineffective_root_store_removals}).

\textit{Intermediate \xscs{}.}
\label{sec:crosssign_certificates:intermediate-xs}
Intermediate \xscs{} contain two or more intermediate CA certificates from different issuers, but no root store certificate (cf. Figure~\ref{fig:crosssign_certificates:xs_types}b).
They are similar to root \xscs{}, but no certificate in the \xsc{} is directly trusted by root store maintainers.
The same benefits (bootstrapping, large trust coverage, and compatibility with old applications) apply.
Additionally, CAs without an own root certificate may employ such cross-signing to root trust in multiple CAs, and thus achieve independence from a single business partner.
However, intermediate \xscs{} also are problematic if certificate revocations are not thoroughly applied to all certificates of a \xsc{} (cf. Section~\ref{sec:crosssigns_in_the_wild:ineffective_revocations}).

\textit{Leaf \xscs{}.}
\label{sec:crosssign_certificates:leaf-xs}
Leaf \xscs{} contain only leaf certificates (cf. Figure~\ref{fig:crosssign_certificates:xs_types}c).
This could in principle be used in cases where some client applications only trust divergent CAs; Here, the server can send \emph{that} certificate of which it knows that it will be trusted by the client, e.g., using the user agent data in an HTTP request to determine the client's root store.
As leaf certificates cannot issue certificates, problematic cases are limited in scope.
Still, for a revocation in case of a security incident, all certificates of the \xsc{} have to be revoked -- otherwise insecure certificates remain valid.

\textit{Leaf-Mix \xscs{} (Theoretical).}
\label{sec:crosssign_certificates:leaf-mix-xs}
In theory, a \xsc{} could contain leaf certificates and (root or intermediate) CA certificates (cf. Figure~\ref{fig:crosssign_certificates:xs_types}d).
Private keys of CA certificates have special protection requirements such as the use of sealed hardware.
As private keys of leaf certificates are typically present on comparably vulnerable end-systems, a leaf-mix \xsc{} would put a key that can issue certificates at high risk.
Fortunately, our dataset does not contain any valid leaf-mix \xscs{}.

\section{Datasets and Methodology}
\def\crtshTotalCrts{\num{156315}}
\def\crtshAlreadyKnownCrts{\num{2635}}
\def\crtshAddedCrts{\num{153680}}
\def\crtshGoogleCTprecertSigning{\num{147439}}
\def\crtshUsefulAddedCrts{\num{6241}}
The analysis in this paper is based on a dataset of certificates used in the wild. The certificates are passively collected from the outgoing SSL/TLS connections on all ports of several universities and research networks mainly located in North America.
Beyond public data (e.g., CT Logs), it also contains certificates that are not publicly available (e.g., user or private certificates) and thereby enables us to take a broad yet unseen perspective on cross-signing.
Our dataset spans a period of more than 7 years, starting February 2012 and ending August 2019, covering more than \num{300} billion TLS connections.
For a broad view on CA certificates, we extend this dataset with CA certificates from CT logs\footnote{These contain \crtshTotalCrts\ CA certificates, however, \crtshGoogleCTprecertSigning\ of them are used for \emph{Google Certificate Transparency (Precert Signing)} which we exclude from our analysis. Further \crtshAlreadyKnownCrts\ are already included in our passive dataset. CT thus adds \crtshUsefulAddedCrts\ CA certificates.} (those logs used by \href{https://crt.sh}{crt.sh} \cite{crt_sh_ct_logs}).

The passive data collection effort was cleared by the respective responsible party at every contributing institution.
Our collection effort focuses on machine to machine communication and excludes or anonymizes sensitive information. See Appendix~\ref{app:data_ethics} for details.

\label{special:reproducibility:toolchain_release}
We release the toolchain~\cite{toolrelease} that enables the analysis of the cross-signing relationships of X.509 certificates from any source (e.g., Censys~\cite{censys15} or CT logs~\cite{crt_sh_ct_logs}).
Since that part of our data set derived from passive measurements is subject to NDAs and contains private certificates that are not contained in public repositories, we are unable to release our data set.
Our toolchain can, however, be used to reproduce our analysis using other data sources, e.g., CT.

We validate certificates using our custom validation logic which closely mirrors the way browsers perform certificate validation.
We build potential paths using name-matching between subject and issuer fields.
Unlike browsers, we build all possible paths from a certificate to root certificates\footnote{Due to computational complexity, we base our evaluation for some CT certificates on all paths of length 12 or smaller (multiple million paths per certificate); for the vast majority of certificates, also longer paths were validated and used.}. We do not just search for the shortest possible path.
To this end, we check each path separately by passing the complete path to OpenSSL for validation and using only the specific root certificate for this path as trust anchor.
We obtain the earliest and latest possible validity dates based on the \emph{not before} and \emph{not after} dates of the path's certificates and use these for the validation.
We also check if a path is valid given the path-length-constraints of its CA certificates.
To obtain validity information for a certificate, all found paths and validity period(s) of the certificate are mapped to the corresponding root store versions---based on the trust status of a path's root certificate in root stores over time.

We use the root stores of major operating systems, web browsers, and grid networks and consider certificate additions as well as removals across their versions.
Specifically, we use the root stores of Microsoft~\cite{rootstore_microsoft}, Apple (iOS, OSX)~\cite{rootstore_apple}, Google (Android)~\cite{rootstore_android}, Mozilla (Firefox, Linux distributions)~\cite{rootstore_mozilla}, and the grid computing PKI~\cite{rootstore_igtf}.
We also include the PKIs of the governments of the United States \cite{rootstore_us}, Australia \cite{rootstore_aus}, Switzerland \cite{rootstore_swiss}, Oman \cite{rootstore_oman}, Netherlands \cite{rootstore_netherland}, Japan \cite{rootstore_jp}, India \cite{rootstore_india}, and Estonia \cite{rootstore_estonia}.

We test for revocations based on browser revocation lists, i.e., Mozilla's OneCRL, Google's CRLSets and blacklist, Apple's lists of blocked certificates, and Microsoft's root store program.
We also use revocation status information from \href{https://crt.sh}{crt.sh} for CA CRLs and Microsoft's \emph{disallowedcert.stl} blacklist to cross-check with the former web browser or operating system revocation mechanisms.

\def\datasetAllCertsCnt{\num{225243355}}
\def\datasetValidLeafCertsCAtrueCnt{\num{23247834}}
\def\datasetValidLeafCertsCAnullCnt{\num{247337}}
\def\datasetValidCertsCnt{\num{23504394}}
\def\datasetCertsWithValidPathCnt{\num{23503674}}
\def\datasetRootcerts{\num{793}}
\def\datasetRootcertsSelfsignedNonCAvOne{\num{23}}
\def\datasetRootcertsSelfsignedNonCAvThreeLegacy{\num{3}}
\def\datasetRootcertsSelfsignedNonCA{\num{26}}
\def\datasetRootcertsSelfsigned{\num{694}}
\def\datasetRootcertsNotSelfsignedCA{\num{73}}
\def\datasetRootcertsNotSelfsignedCAWithPath{\num{73}}
\def\datasetRootcertsWithoutPath{\num{720}}
\def\datasetNonCAcertsWithValidPath{\num{23495171}}
\def\datasetValidCAcerts{\num{9197}}
Our dataset includes \datasetAllCertsCnt\ certificates of which \datasetValidCertsCnt\ are valid\footnote{
\crtshGoogleCTprecertSigning\ invalid certificates are for \emph{Google Certificate Transparency Precert Signing} (excluded from analysis).
Most other invalid certificates are a side effect of observing real Internet traffic:
70\% are used for user authentication in the grid and not trusted by CAs.
Auto-generated certificates for Tor or WebRTC account for 17\% and 1\%, respectively.
9\% are self-signed.
The remaining 3\% comprise IoT and firewall certificates.
Earlier studies already found invalid certificates to be common in the web~\cite{Chung2016_invalid}.
}
(\datasetRootcertsWithoutPath\ only due to inclusion in a root store; \datasetCertsWithValidPathCnt\ have a valid path to a root store).
We focus our analysis on these valid certificates. They split into \datasetValidCAcerts\ CA certificates, \datasetRootcertsSelfsignedNonCAvOne\ self-signed X509v1 certificates, \datasetRootcertsSelfsignedNonCAvThreeLegacy\ self-signed X509v3 certificates using legacy extensions to show their CA status, and \datasetNonCAcertsWithValidPath\ leaf certificates. %
\datasetRootcerts\ certificates are included in at least one of our used root stores (all \datasetRootcertsSelfsigned\ self-signed CA certificates, the \datasetRootcertsSelfsignedNonCAvOne\ self-signed X509v1 and \datasetRootcertsSelfsignedNonCAvThreeLegacy\ X509v3 legacy certificates, and \datasetRootcertsNotSelfsignedCA\ not self-signed CA certificates; All the latter \datasetRootcertsNotSelfsignedCAWithPath\ certificates are already valid due to their root store inclusion but additionally provide paths to other trusted roots).

\def\datasetValidPaths{\num{9.3} billion}
\def\datasetLongestPath{\num{17}}
\def\datasetRootcertsThatIssuedCnt{\num{526}}

\def\datasetCertMaxValidationPathCnt{\num{45225135}}
\def\datasetCertMaxValidationPathsCertcnt{\num{1}}
\def\datasetCertMaxValidationPathsCertCertificate{\emph{IdenTrust Global Common Root CA 1}}
\def\datasetLongestPathCerts{\num{63}}

\def\datasetAddtrustExternalCaRootPathCnt{\num{3.7} billion} %

\def\datasetDSTrootCAxThreeCerts{\num{8473760}}
\def\datasetISRGrootXoneCerts{\num{8382825}}
\def\datasetAddtrustExternalCaRootCerts{\num{5966846}}

To give a short overview of our dataset:
We see \datasetRootcertsThatIssuedCnt\ of the \datasetRootcerts\ root certificates being used---meaning that we found another certificate that was signed by them in our dataset.
In total, we find more than \datasetValidPaths\ valid paths between the certificates in our dataset. The longest path is \datasetLongestPath\ certificates long (including root and leaf).
We find \datasetLongestPathCerts\ certificates with this path length.
In the entirety of our measurement, the root CA against which we can validate the most certificates was the IdenTrust \emph{DST Root CA X3} (with \datasetDSTrootCAxThreeCerts\ certificates).
This is probably driven by their cross-sign of Let's Encrypt which we will detail later in this paper. The Let's Encrypt \emph{ISRG Root X1} follows as a close second with
\datasetISRGrootXoneCerts\ certificates.
The CA to which we can find most unique paths is the Comodo \emph{AddTrust External CA Root} CA. For this root, we can find \datasetAddtrustExternalCaRootPathCnt\ different validation paths for a total of \datasetAddtrustExternalCaRootCerts\ distinct certificates.
The maximum number of different validation paths (as in different paths through the CA ecosystem through which we could validate it) for a CA certificate we found was \datasetCertMaxValidationPathCnt.
These paths provide validity for the intermediate \datasetCertMaxValidationPathsCertCertificate.

\def\csLeafNum{\num{47221}}
\def\reissuanceLeafNum{\num{97233}}
\def\csTotalNum{\num{47543}} %
\def\reissuanceTotalNum{\num{97240}} %
\def\csRootNum{\num{86}} %
\def\reissuanceRootNum{\num{1}}
  \def\csRootInternOnlyNum{\num{58}} %
  \def\csRootExternNum{\num{28}} %
    \def\csRootExternSingleNum{\num{16}} %
    \def\csRootExternMultiNum{\num{12}}
\def\csIntermediateNum{\num{236}}
\def\reissuanceIntermediateNum{\num{6}}
  \def\csIntermediateIntern{\num{70}}
  \def\csIntermediateExtern{\num{166}}
    \def\csIntermediateExternWithIntern{\num{42}}
    \def\csIntermediateExternNoIntern{\num{124}}
    \def\csIntermediateExternMultiCANum{\num{149}}

\def\leafXSpercentDigicert{67\%}
\def\leafXSpercentSymantecOfDigicertCases{45\%}
\def\leafXSpercentGoDaddyCases{10\%}
\def\leafXSpercentComodoCases{8\%}
\def\leafXSpercentLECases{significantly less than 1\%}
\def\leafXSpercentOneCA{63\%}
\def\leafXSpercentMutliCAs{37\%}

Taking a look at cross-signs, we find a total of \csTotalNum\ \xscs{}.
These split into \csRootNum\ root, \csIntermediateNum\ intermediate, and \csLeafNum\ leaf \xscs{}\footnote{The number of leaf \xscs{} is solely based on our passive dataset.}.
For the remainder of this paper, we focus our analysis on root and intermediate CA certificates as these cross-signs impact the trust of thousands or millions of leaf-certificates.
We however note that it is interesting that there is a significant number of leaf %
\xscs{} which may bring along the same problems as cross-signed CA certificates, e.g., incomplete revocation.
We note that CRLite~\cite{larisch2017_crlite} (now being integrated in Firefox) potentially enables large scale revocation for leaf certificates---by reducing the overhead for CRL updates with efficient Bloom filter-based incremental updates.

\label{special:leafXS_basic_information}
No clear phenomenon explains the leaf \xscs{}: For \leafXSpercentOneCA\ of leaf \xscs{}, one CA issued all certificates (multiple CAs: \leafXSpercentMutliCAs). DigiCert is involved in \leafXSpercentDigicert\ of leaf \xscs{} (\leafXSpercentSymantecOfDigicertCases\ of those stem from Symantec before DigiCert acquired it), followed by GoDaddy (\leafXSpercentGoDaddyCases) and Comodo (\leafXSpercentComodoCases). Let's Encrypt, however, is involved in only \leafXSpercentLECases\ of cases.
In future work it might be worth trying to explore the motivation behind these cases.

When analyzing the root and intermediate \xscs{}, we find \emph{internal} and \emph{external cross-signs}.
Internal \xscs{} comprise only certificates that have been issued within the same CA group, i.e., issuer and certificate owner are controlled by the same entity.
For example, we find the two largest CA groups, Comodo\footnote{By \emph{Comodo}, we refer to the CA business, which was recently renamed to \emph{Sectigo} to avoid naming confusion with the separate company \emph{Comodo Group} \cite{comodo_rebranding_announcement}} and DigiCert, to internally cross-sign certificates within and across their subsidiary CAs, both considered internal cross-signing.

External \xscs{} cross boundaries of CA groups, i.e., issuer and certificate owner of at least one certificate of the \xsc{} are controlled by different entities.
These are particularly interesting as issuing CAs take responsibility for the actions of the resulting intermediates \cite{mozilla_roostorepolicy}; whereas the information flow between organizations may be limited, challenging fast and thorough revocation.

We find internal and external cross-signing for both, root (\csRootInternOnlyNum\ internal / \csRootExternNum\ external) as well as intermediate (\csIntermediateIntern\ / \csIntermediateExtern) \xscs{}.
For intermediate CA certificates, we find \csIntermediateExternNoIntern\ cases where the organization does not have their own root certificate. These organizations are completely dependent on the cross-signs from their issuers.

In addition to these \xscs{}, we also identify \reissuanceTotalNum\ cases of \emph{certificate re-issuances}.
Certificate re-issuances are similar to cross-signing: there are several certificates using the same subject and key.
The difference to \xscs{} is that the validity periods of these certificates overlap only slightly or not at all (note that key-reuse across certificates with \emph{different} subjects clearly distinguishes from cross-signing).
We only consider cases as \xscs{} if their certificate's validities overlap for \num{121} or more days (cf. Section~\ref{special:xs_validity_overlap_min}) and exclude re-issuances from our following discussion.
Notably, certificate re-issuing is prevalent for leaf certificates (\reissuanceLeafNum) and almost non-existent when root (\reissuanceRootNum) or intermediate (\reissuanceIntermediateNum) certificates are involved.
We still analyzed the latter, but found them uninteresting.

\section{Cross-Signs in the Wild} %
\label{sec:crosssigns_in_the_wild}

\def\DBcntBootstrappingExternal{57}  %
\def\DBcntBootstrappingInternal{\fix{\textbf{XX}}}

\def\DBcntExpandingTrustStores{64}

\def\DBcntExpandingTrustTime{46}

\def\DBcntAlternativePaths{155}  %

\def\DBcntMultipleSignatureAlgorithms{23}

\def\manualcountValidDespiteRevocation{16}  %
\def\manualcountPKIbarrierBreaches{7}  %
\def\manualcountOwnershipChange{35}  %
\def\manualcountBackdating{7}  %
\def\manualcountMissingTransparency{2}  %

\begin{table}
  \caption{Observed \xscs{} by category. Numbers marked with $\ast$ are based on manual investigation of the passive dataset. A \xsc{} can count to multiple categories.}
  \label{tab:crosssigns_in_the_wild:quantitative_overview}
  \begin{tabular}{clrl}
    \toprule
      \symbolBadXS      & Valid after revocation                    & \manualcountValidDespiteRevocation    & $\ast$ \\
      \symbolBadXS      & PKI barrier breaches                      & \manualcountPKIbarrierBreaches        & $\ast$ \\
      \symbolGoodXS     & Bootstrapping %
                                                                    & \DBcntBootstrappingExternal           & \\
      \symbolNeutralXS  & Expanded trust (new stores / longer time) %
                                                                    & \DBcntExpandingTrustStores\ / \DBcntExpandingTrustTime  & \\
      \symbolNeutralXS  & Alternative paths                         & \DBcntAlternativePaths                & \\
      \symbolNeutralXS  & Support of multiple signature algorithms  & \DBcntMultipleSignatureAlgorithms     & \\
      \symbolNeutralXS  & Ownership change                          & \manualcountOwnershipChange           & $\ast$ \\
      \symbolUglyXS     & Backdating                                & \manualcountBackdating                & $\ast$ \\
      \symbolUglyXS     & Missing transparency                      & \manualcountMissingTransparency       & $\ast$ \\
    \bottomrule
\end{tabular}
\end{table}

In this section, we systematically analyze the use of cross-signing in existing PKIs.
To identify the different categories of cross-signing and their motivation, we start off from its primary goal: extending certificate trust---exploiting all derived trust paths in our dataset.
We combine the derived validity status in root stores (over time) with revocation data and enrich this with knowledge on ownership and (governmental-)control over CAs, and algorithmic properties of certificates.
We overview the identified categories in Table~\ref{tab:crosssigns_in_the_wild:quantitative_overview} with their number of occurrence in our dataset(s).
We focus on
(i) the bad~\symbolBadXS\hspace{1pt}: potential and exploited security problems
(Sections~\ref{sec:crosssigns_in_the_wild:ineffective_revocations} - \ref{sec:crosssigns_in_the_wild:pki_barrier_breaches}),
(ii) the good~\symbolGoodXS\hspace{1pt}: ease bootstrapping of new CAs (Section~\ref{sec:crosssigns_in_the_wild:bootstrapping_help}),
(iii) cases with pros and cons~[\symbolNeutralXS], e.g., the transition to new cryptographic algorithms (Sections~\ref{sec:crosssigns_in_the_wild:trust_coverage_and_fallback_trust} - \ref{sec:crosssigns_in_the_wild:ownership_changes}), and
(iv) the ugly~\symbolUglyXS\ behavior due to problematic practices and lacking transparency (Section~\ref{sec:crosssigns_in_the_wild:misc}).

\subsection{Valid Paths After Revocations} %
\label{sec:crosssigns_in_the_wild:ineffective_revocations}
\label{sec:crossigns_problems:ineffective_root_store_removals}

Cross-signs significantly complicate the revocation of CA certificates.
When a CA certificate is revoked, \emph{all} of its cross-signs need to be revoked as well.
A single remaining unrevoked cross-sign means that a valid path to a trusted root still exists---and all certificates of the revoked CA can still be validated.
As we will show in this section, cross-signing repeatedly caused incomplete revocations, or even added new valid trust paths for already revoked certificates.
We next empirically study this problem of \emph{valid paths after revocation} that adds an additional layer of complexity to the already fragile revocation mechanisms in PKIs.
We focus our discussion on cross-signs that caused incomplete revocations for WoSign and DigiNotar as well as incomplete revocations in vendor-controlled CRLs (especially Mozillas' OneCRL and Googles' CRLSet).

\subsubsection{The WoSign and StartCom Ban} %
\begin{figure}
    \centering

\begin{tikzpicture}
\usetikzlibrary{positioning}
\usetikzlibrary{patterns}

\draw[line width=1pt] (0,0) -- (0.97*0.86\columnwidth, 0);
\node[] (2016) at (0*0.97*0.86\columnwidth/3 + 0*0.97*0.86\columnwidth/3/12 + 0*0.97*0.86\columnwidth/3/12/31,-8pt) {\tiny \textbf{2016}};
\draw (0*0.97*0.86\columnwidth/3 + 0*0.97*0.86\columnwidth/3/12 + 0*0.97*0.86\columnwidth/3/12/31,0) -- (2016);
\node[] (2017) at (1*0.97*0.86\columnwidth/3 + 0*0.97*0.86\columnwidth/3/12 + 0*0.97*0.86\columnwidth/3/12/31,-8pt) {\tiny \textbf{2017}};
\draw (1*0.97*0.86\columnwidth/3 + 0*0.97*0.86\columnwidth/3/12 + 0*0.97*0.86\columnwidth/3/12/31,0) -- (2017);
\node[] (2018) at (2*0.97*0.86\columnwidth/3 + 0*0.97*0.86\columnwidth/3/12 + 0*0.97*0.86\columnwidth/3/12/31,-8pt) {\tiny \textbf{2018}};
\draw (2*0.97*0.86\columnwidth/3 + 0*0.97*0.86\columnwidth/3/12 + 0*0.97*0.86\columnwidth/3/12/31,0) -- (2018);
\node[] (2019) at (3*0.97*0.86\columnwidth/3 + 0*0.97*0.86\columnwidth/3/12 + 0*0.97*0.86\columnwidth/3/12/31,-8pt) {\tiny \textbf{2019}};
\draw (3*0.97*0.86\columnwidth/3 + 0*0.97*0.86\columnwidth/3/12 + 0*0.97*0.86\columnwidth/3/12/31,0) -- (2019);

\node[] (ev1_anchor_timeline) at (2*0.97*0.86\columnwidth/3 + 7*0.97*0.86\columnwidth/3/12 + 14*0.97*0.86\columnwidth/3/12/31,0) {};
\node[above = 50pt of ev1_anchor_timeline.center] (ev1_anchor_text) {};
\node[above = -3pt of ev1_anchor_text.center] (ev1_label) {\scriptsize \textbf{WoSign/StartCom ...}};

\node[] (ev2_anchor_timeline) at (2*0.97*0.86\columnwidth/3 + 8*0.97*0.86\columnwidth/3/12 + 7*0.97*0.86\columnwidth/3/12/31,0) {};
\node[above = 38pt of ev2_anchor_timeline.center] (ev2_anchor_text) {};
\node[above = -3pt of ev2_anchor_text.center] (ev2_label) {\scriptsize \textbf{\normalfont not trusted in {\it store}}};

\node[] (ev3_anchor_timeline) at (2*0.97*0.86\columnwidth/3 + 3*0.97*0.86\columnwidth/3/12 + 8*0.97*0.86\columnwidth/3/12/31,0) {};
\node[above = 39pt of ev3_anchor_timeline.center] (ev3_anchor_text) {};
\node[above = -3pt of ev3_anchor_text.center] (ev3_label) {\scriptsize \textbf{\normalfont {\normalfont\tiny\faCalendarTimesO}}};

\node[] (ev4_anchor_timeline) at (2*0.97*0.86\columnwidth/3 + 8*0.97*0.86\columnwidth/3/12 + 17*0.97*0.86\columnwidth/3/12/31,0) {};
\node[above = 43pt of ev4_anchor_timeline.center] (ev4_anchor_text) {};
\node[above = -3pt of ev4_anchor_text.center] (ev4_label) {\scriptsize \textbf{\normalfont certs issued {\it later} are}};

\node[] (ev5_anchor_timeline) at (2*0.97*0.86\columnwidth/3 + 8*0.97*0.86\columnwidth/3/12 + 0*0.97*0.86\columnwidth/3/12/31,0) {};
\node[above = 29pt of ev5_anchor_timeline.center] (ev5_anchor_text) {};
\node[above = -3pt of ev5_anchor_text.center] (ev5_label) {\scriptsize \textbf{\normalfont valid path via XS*}};

\node[] (ev6_anchor_timeline) at (2*0.97*0.86\columnwidth/3 + 3*0.97*0.86\columnwidth/3/12 + 7*0.97*0.86\columnwidth/3/12/31,0) {};
\node[above = 22pt of ev6_anchor_timeline.center] (ev6_anchor_text) {};
\node[above = -3pt of ev6_anchor_text.center] (ev6_label) {\scriptsize \textbf{\normalfont {\normalfont\tiny\faMinusCircle}}};

\node[] (ev7_anchor_timeline) at (2*0.97*0.86\columnwidth/3 + 8*0.97*0.86\columnwidth/3/12 + 10*0.97*0.86\columnwidth/3/12/31,0) {};
\node[above = 22pt of ev7_anchor_timeline.center] (ev7_anchor_text) {};
\node[above = -3pt of ev7_anchor_text.center] (ev7_label) {\scriptsize \textbf{\normalfont removed from {\it store}}};

\node[] (ev8_bg_anchor_timeline) at (1*0.97*0.86\columnwidth/3 + 8*0.97*0.86\columnwidth/3/12 + 25*0.97*0.86\columnwidth/3/12/31,0) {};
\node[above = 44pt of ev8_bg_anchor_timeline.center] (ev8_bg_anchor_text) {};
\node[fill=white, text opacity=0, above = -3pt of ev8_bg_anchor_text.center] (ev8_bg_date) {\tiny 26 Sep};
\node[fill=white, text opacity=0, above = -4pt of ev8_bg_date] (ev8_bg_label) {\scriptsize \textbf{{\normalfont\tiny\faCalendarTimesO} Windows 10}};
\draw (ev8_bg_anchor_timeline.center) -- (ev8_bg_date);

\node[] (ev8_anchor_timeline) at (1*0.97*0.86\columnwidth/3 + 8*0.97*0.86\columnwidth/3/12 + 25*0.97*0.86\columnwidth/3/12/31,0) {};
\node[above = 44pt of ev8_anchor_timeline.center] (ev8_anchor_text) {};
\node[above = -3pt of ev8_anchor_text.center] (ev8_date) {\tiny 26 Sep};
\node[above = -4pt of ev8_date] (ev8_label) {\scriptsize \textbf{{\normalfont\tiny\faCalendarTimesO} Windows 10}};

\node[] (ev9_bg_anchor_timeline) at (0*0.97*0.86\columnwidth/3 + 11*0.97*0.86\columnwidth/3/12 + 0*0.97*0.86\columnwidth/3/12/31,0) {};
\node[above = 33pt of ev9_bg_anchor_timeline.center] (ev9_bg_anchor_text) {};
\node[fill=white, text opacity=0, above = -3pt of ev9_bg_anchor_text.center] (ev9_bg_date) {\tiny 1 Dec};
\node[fill=white, text opacity=0, above = -4pt of ev9_bg_date] (ev9_bg_label) {\scriptsize \textbf{{\normalfont\tiny\faCalendarTimesO} Apple}};
\draw (ev9_bg_anchor_timeline.center) -- (ev9_bg_date);

\node[] (ev9_anchor_timeline) at (0*0.97*0.86\columnwidth/3 + 11*0.97*0.86\columnwidth/3/12 + 0*0.97*0.86\columnwidth/3/12/31,0) {};
\node[above = 33pt of ev9_anchor_timeline.center] (ev9_anchor_text) {};
\node[above = -3pt of ev9_anchor_text.center] (ev9_date) {\tiny 1 Dec};
\node[above = -4pt of ev9_date] (ev9_label) {\scriptsize \textbf{{\normalfont\tiny\faCalendarTimesO} Apple}};

\node[] (ev10_bg_anchor_timeline) at (0*0.97*0.86\columnwidth/3 + 0*0.97*0.86\columnwidth/3/12 + 0*0.97*0.86\columnwidth/3/12/31,0) {};
\node[above = 22pt of ev10_bg_anchor_timeline.center] (ev10_bg_anchor_text) {};
\node[fill=white, text opacity=0, above = -3pt of ev10_bg_anchor_text.center] (ev10_bg_date) {\tiny 1 Jan};
\node[fill=white, text opacity=0, above = -4pt of ev10_bg_date] (ev10_bg_label) {\scriptsize \textbf{SHA1 deprecated}};
\draw (ev10_bg_anchor_timeline.center) -- (ev10_bg_date);

\node[] (ev10_anchor_timeline) at (0*0.97*0.86\columnwidth/3 + 0*0.97*0.86\columnwidth/3/12 + 0*0.97*0.86\columnwidth/3/12/31,0) {};
\node[above = 22pt of ev10_anchor_timeline.center] (ev10_anchor_text) {};
\node[above = -3pt of ev10_anchor_text.center] (ev10_date) {\tiny 1 Jan};
\node[above = -4pt of ev10_date] (ev10_label) {\scriptsize \textbf{SHA1 deprecated}};

\node[] (ev11_bg_anchor_timeline) at (1*0.97*0.86\columnwidth/3 + 8*0.97*0.86\columnwidth/3/12 + 0*0.97*0.86\columnwidth/3/12/31,0) {};
\node[above = 22pt of ev11_bg_anchor_timeline.center] (ev11_bg_anchor_text) {};
\node[fill=white, text opacity=0, above = -3pt of ev11_bg_anchor_text.center] (ev11_bg_date) {\tiny Sep};
\node[fill=white, text opacity=0, above = -4pt of ev11_bg_date] (ev11_bg_label) {\scriptsize \textbf{{\normalfont\tiny\faMinusCircle} Google}};
\draw (ev11_bg_anchor_timeline.center) -- (ev11_bg_date);

\node[] (ev11_anchor_timeline) at (1*0.97*0.86\columnwidth/3 + 8*0.97*0.86\columnwidth/3/12 + 0*0.97*0.86\columnwidth/3/12/31,0) {};
\node[above = 22pt of ev11_anchor_timeline.center] (ev11_anchor_text) {};
\node[above = -3pt of ev11_anchor_text.center] (ev11_date) {\tiny Sep};
\node[above = -4pt of ev11_date] (ev11_label) {\scriptsize \textbf{{\normalfont\tiny\faMinusCircle} Google}};

\node[] (ev12_bg_anchor_timeline) at (0*0.97*0.86\columnwidth/3 + 9*0.97*0.86\columnwidth/3/12 + 20*0.97*0.86\columnwidth/3/12/31,0) {};
\node[above = 3pt of ev12_bg_anchor_timeline.center] (ev12_bg_anchor_text) {};
\node[fill=white, text opacity=0, above = -3pt of ev12_bg_anchor_text.center] (ev12_bg_date) {\tiny 21 Oct};
\node[fill=white, text opacity=0, above = -4pt of ev12_bg_date] (ev12_bg_label) {\scriptsize \textbf{{\normalfont\tiny\faCalendarTimesO} Mozilla/Google}};
\draw (ev12_bg_anchor_timeline.center) -- (ev12_bg_date);

\node[] (ev12_anchor_timeline) at (0*0.97*0.86\columnwidth/3 + 9*0.97*0.86\columnwidth/3/12 + 20*0.97*0.86\columnwidth/3/12/31,0) {};
\node[above = 3pt of ev12_anchor_timeline.center] (ev12_anchor_text) {};
\node[above = -3pt of ev12_anchor_text.center] (ev12_date) {\tiny 21 Oct};
\node[above = -4pt of ev12_date] (ev12_label) {\scriptsize \textbf{{\normalfont\tiny\faCalendarTimesO} Mozilla/Google}};

\node[] (ev13_bg_anchor_timeline) at (2*0.97*0.86\columnwidth/3 + 0*0.97*0.86\columnwidth/3/12 + 0*0.97*0.86\columnwidth/3/12/31,0) {};
\node[above = 3pt of ev13_bg_anchor_timeline.center] (ev13_bg_anchor_text) {};
\node[fill=white, text opacity=0, above = -3pt of ev13_bg_anchor_text.center] (ev13_bg_date) {\tiny Jan};
\node[fill=white, text opacity=0, above = -4pt of ev13_bg_date] (ev13_bg_label) {\scriptsize \textbf{{\normalfont\tiny\faMinusCircle} Mozilla}};
\draw (ev13_bg_anchor_timeline.center) -- (ev13_bg_date);

\node[] (ev13_anchor_timeline) at (2*0.97*0.86\columnwidth/3 + 0*0.97*0.86\columnwidth/3/12 + 0*0.97*0.86\columnwidth/3/12/31,0) {};
\node[above = 3pt of ev13_anchor_timeline.center] (ev13_anchor_text) {};
\node[above = -3pt of ev13_anchor_text.center] (ev13_date) {\tiny Jan};
\node[above = -4pt of ev13_date] (ev13_label) {\scriptsize \textbf{{\normalfont\tiny\faMinusCircle} Mozilla}};

\node[] (ev14_bg_anchor_timeline) at (0*0.97*0.86\columnwidth/3 + 9*0.97*0.86\columnwidth/3/12 + 3*0.97*0.86\columnwidth/3/12/31,0) {};
\node[below = 45pt of ev14_bg_anchor_timeline.center] (ev14_bg_anchor_text) {};
\node[fill=white, text opacity=0, below = -3pt of ev14_bg_anchor_text.center] (ev14_bg_date) {\tiny 4 Oct};
\node[fill=white, text opacity=0, below = -4pt of ev14_bg_date] (ev14_bg_label) {\scriptsize \textbf{Comodo revokes XSs}};
\draw (ev14_bg_anchor_timeline.center) -- (ev14_bg_date);

\node[] (ev14_anchor_timeline) at (0*0.97*0.86\columnwidth/3 + 9*0.97*0.86\columnwidth/3/12 + 3*0.97*0.86\columnwidth/3/12/31,0) {};
\node[below = 45pt of ev14_anchor_timeline.center] (ev14_anchor_text) {};
\node[below = -3pt of ev14_anchor_text.center] (ev14_date) {\tiny 4 Oct};
\node[below = -4pt of ev14_date] (ev14_label) {\scriptsize \textbf{Comodo revokes XSs}};

\node[] (ev15_bg_anchor_timeline) at (1*0.97*0.86\columnwidth/3 + 9*0.97*0.86\columnwidth/3/12 + 17*0.97*0.86\columnwidth/3/12/31,0) {};
\node[below = 45pt of ev15_bg_anchor_timeline.center] (ev15_bg_anchor_text) {};
\node[fill=white, text opacity=0, below = -3pt of ev15_bg_anchor_text.center] (ev15_bg_date) {\tiny 18 Oct};
\node[fill=white, text opacity=0, below = -4pt of ev15_bg_date] (ev15_bg_label) {\scriptsize \textbf{Certinomis revokes XS*}};
\draw (ev15_bg_anchor_timeline.center) -- (ev15_bg_date);

\node[] (ev15_anchor_timeline) at (1*0.97*0.86\columnwidth/3 + 9*0.97*0.86\columnwidth/3/12 + 17*0.97*0.86\columnwidth/3/12/31,0) {};
\node[below = 45pt of ev15_anchor_timeline.center] (ev15_anchor_text) {};
\node[below = -3pt of ev15_anchor_text.center] (ev15_date) {\tiny 18 Oct};
\node[below = -4pt of ev15_date] (ev15_label) {\scriptsize \textbf{Certinomis revokes XS*}};

\node[] (ev16_bg_anchor_timeline) at (2*0.97*0.86\columnwidth/3 + 9*0.97*0.86\columnwidth/3/12 + 14*0.97*0.86\columnwidth/3/12/31,0) {};
\node[below = 35pt of ev16_bg_anchor_timeline.center] (ev16_bg_anchor_text) {};
\node[fill=white, text opacity=0, below = -3pt of ev16_bg_anchor_text.center] (ev16_bg_label) {\scriptsize \textbf{\bf mid 2019\guillemotright}};
\node[fill=white, text opacity=0, below = -4pt of ev16_bg_label] (ev16_bg_sublabel) {\tiny \bf\scriptsize Certinomis {\normalfont\tiny\faMinusCircle}};

\node[] (ev16_anchor_timeline) at (2*0.97*0.86\columnwidth/3 + 9*0.97*0.86\columnwidth/3/12 + 14*0.97*0.86\columnwidth/3/12/31,0) {};
\node[below = 35pt of ev16_anchor_timeline.center] (ev16_anchor_text) {};
\node[below = -3pt of ev16_anchor_text.center] (ev16_label) {\scriptsize \textbf{\bf mid 2019\guillemotright}};
\node[below = -4pt of ev16_label] (ev16_sublabel) {\tiny \bf\scriptsize Certinomis {\normalfont\tiny\faMinusCircle}};

\node[] (ev17_bg_anchor_timeline) at (1*0.97*0.86\columnwidth/3 + 8*0.97*0.86\columnwidth/3/12 + 0*0.97*0.86\columnwidth/3/12/31,0) {};
\node[below = 25pt of ev17_bg_anchor_timeline.center] (ev17_bg_anchor_text) {};
\node[fill=white, text opacity=0, below = -3pt of ev17_bg_anchor_text.center] (ev17_bg_date) {\tiny Sep};
\node[fill=white, text opacity=0, below = -4pt of ev17_bg_date] (ev17_bg_label) {\scriptsize \textbf{OneCRL/CRLSet add XS*}};
\draw (ev17_bg_anchor_timeline.center) -- (ev17_bg_date);

\node[] (ev17_anchor_timeline) at (1*0.97*0.86\columnwidth/3 + 8*0.97*0.86\columnwidth/3/12 + 0*0.97*0.86\columnwidth/3/12/31,0) {};
\node[below = 25pt of ev17_anchor_timeline.center] (ev17_anchor_text) {};
\node[below = -3pt of ev17_anchor_text.center] (ev17_date) {\tiny Sep};
\node[below = -4pt of ev17_date] (ev17_label) {\scriptsize \textbf{OneCRL/CRLSet add XS*}};

\node[] (ev18_bg_anchor_timeline) at (2*0.97*0.86\columnwidth/3 + 10*0.97*0.86\columnwidth/3/12 + 0*0.97*0.86\columnwidth/3/12/31,0) {};
\node[below = 15pt of ev18_bg_anchor_timeline.center] (ev18_bg_anchor_text) {};
\node[fill=white, text opacity=0, below = -3pt of ev18_bg_anchor_text.center] (ev18_bg_label) {\scriptsize \textbf{\bf 2027\guillemotright}};
\node[fill=white, text opacity=0, below = -4pt of ev18_bg_label] (ev18_bg_sublabel) {\tiny \bf\scriptsize XS* expiry};

\node[] (ev18_anchor_timeline) at (2*0.97*0.86\columnwidth/3 + 10*0.97*0.86\columnwidth/3/12 + 0*0.97*0.86\columnwidth/3/12/31,0) {};
\node[below = 15pt of ev18_anchor_timeline.center] (ev18_anchor_text) {};
\node[below = -3pt of ev18_anchor_text.center] (ev18_label) {\scriptsize \textbf{\bf 2027\guillemotright}};
\node[below = -4pt of ev18_label] (ev18_sublabel) {\tiny \bf\scriptsize XS* expiry};

\node[] (ev19_bg_anchor_timeline) at (1*0.97*0.86\columnwidth/3 + 3*0.97*0.86\columnwidth/3/12 + 12*0.97*0.86\columnwidth/3/12/31,0) {};
\node[below = 10pt of ev19_bg_anchor_timeline.center] (ev19_bg_anchor_text) {};
\node[fill=white, text opacity=0, below = -3pt of ev19_bg_anchor_text.center] (ev19_bg_date) {\tiny 13 Apr};
\node[fill=white, text opacity=0, below = -4pt of ev19_bg_date] (ev19_bg_label) {\scriptsize \textbf{Certinomis: XS*}};
\draw (ev19_bg_anchor_timeline.center) -- (ev19_bg_date);

\node[] (ev19_anchor_timeline) at (1*0.97*0.86\columnwidth/3 + 3*0.97*0.86\columnwidth/3/12 + 12*0.97*0.86\columnwidth/3/12/31,0) {};
\node[below = 10pt of ev19_anchor_timeline.center] (ev19_anchor_text) {};
\node[below = -3pt of ev19_anchor_text.center] (ev19_date) {\tiny 13 Apr};
\node[below = -4pt of ev19_date] (ev19_label) {\scriptsize \textbf{Certinomis: XS*}};

\fill[color=red] (1*0.97*0.86\columnwidth/3 + 3*0.97*0.86\columnwidth/3/12 + 12*0.97*0.86\columnwidth/3/12/31,-3pt) rectangle (1*0.97*0.86\columnwidth/3 + 8*0.97*0.86\columnwidth/3/12 + 0*0.97*0.86\columnwidth/3/12/31,3pt);
\fill[pattern=crosshatch, pattern color=black] (1*0.97*0.86\columnwidth/3 + 3*0.97*0.86\columnwidth/3/12 + 12*0.97*0.86\columnwidth/3/12/31,-3pt) rectangle (1*0.97*0.86\columnwidth/3 + 8*0.97*0.86\columnwidth/3/12 + 0*0.97*0.86\columnwidth/3/12/31,3pt);
\fill[color=red] (2*0.97*0.86\columnwidth/3 + 2*0.97*0.86\columnwidth/3/12 + 24*0.97*0.86\columnwidth/3/12/31,33.5pt) rectangle (2*0.97*0.86\columnwidth/3 + 3*0.97*0.86\columnwidth/3/12 + 20*0.97*0.86\columnwidth/3/12/31,37.5pt);
\fill[pattern=crosshatch, pattern color=black] (2*0.97*0.86\columnwidth/3 + 2*0.97*0.86\columnwidth/3/12 + 24*0.97*0.86\columnwidth/3/12/31,33.5pt) rectangle (2*0.97*0.86\columnwidth/3 + 3*0.97*0.86\columnwidth/3/12 + 20*0.97*0.86\columnwidth/3/12/31,37.5pt);
\draw[draw=black] (2*0.97*0.86\columnwidth/3 + 2*0.97*0.86\columnwidth/3/12 + 14*0.97*0.86\columnwidth/3/12/31,23pt) rectangle (3*0.97*0.86\columnwidth/3 + 1*0.97*0.86\columnwidth/3/12 + 14*0.97*0.86\columnwidth/3/12/31,60pt);

\end{tikzpicture}
    \caption{Certinomis' cross-sign of WoSign/StartCom~(XS*) bypassed \emph{not before} rules ({\hspace{0.3pt}\normalfont\small\faCalendarTimesO}) set up by major root stores.}
    \label{fig:wosign-timeline}
    \Description[The figure shows a timeline from 2016 to 2019.]{On January 1, 2016, SHA1 was deprecated. On October 4, 2016, Comodo revoked its cross-signs of WoSign certificates. On October 21, 2016, Mozilla and Google set up not before rules for WoSign and StartCom. Likewise did Apple on December 1 in the same year. On April 13, 2017, Certinomis cross-signed StartCom (XS*). Certificates obtained a valid path via XS* up to September 2017, when XS* was distrusted by entries in OneCRL, CRLSet. At the same time, Google completely removed WoSign and StartCom roots from its root store. On September 26, 2017, Microsoft set up a not before rule for WoSign and StartCom, but only for Windows 10. On October 18, 2017, Certinomis revoked XS*. In January 2018, Mozilla removed WoSign and StartCom certificates from its root store. According to it's validity period (not after date), the XS* remains valid until 2027. Certinomis was distrusted by root stores mid 2019}
\end{figure}
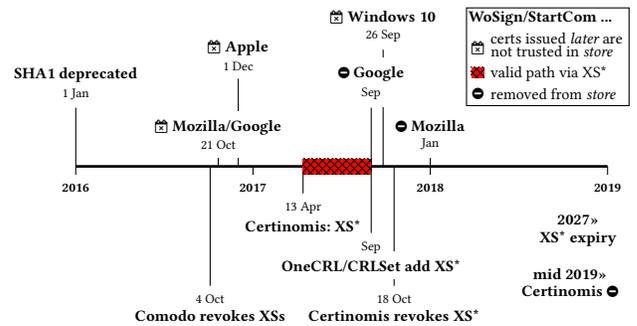

\label{sec:crosssigns_in_the_wild:ineffective_revocations:wosign}
\label{par:crosssigns_in_the_wild:root_cs:external_root_xscs:wosign}
WoSign was distrusted by major root store holders after a series of misbehaviors \cite{mozilla_wosign_incidents}
(see Figure~\ref{fig:wosign-timeline} for a timeline of events).
WoSign did not announce the acquisition of the StartCom CA in time.
It also evaded rules for distrusting SHA1-signed certificates issued after January 1st 2016 by backdating the \emph{not before} date\footnote{The certificate's not before date defines from which time on the certificate is valid, but it is also used as an indicator for the issuance date of the certificate \cite{mozilla_problematic_practices}.} of certificates they issued.
\label{special:not_before_rule}
As a result, Mozilla set up a special \emph{not before rule}, i.e., for certificates that were issued after October 21, 2016, Mozilla distrusted paths that end in a WoSign or StartCom root \cite{mozilla_wosign_distrust_new_certs}.
In January 2018, Mozilla completely removed these roots \cite{mozilla_wosign_distrust_new_certs,mozilla_wosign_removal}.
Google performed similar actions, removing WoSign and StartCom roots around September 2017 \cite{google_wosign_distrust_new_certs,google_wosign_removal}, as did
Apple and Microsoft \cite{apple_wosign_distrust_new_certs,microsoft_wosign_removal}.

By analyzing cross-signs, we observe that WoSign certificates were cross-signed by Comodo subsidiaries, Certplus, Unizeto Certum, and StartCom.
Furthermore, in April 2017---after Google and Mozilla set up \emph{not before} rules for certificates issued by WoSign and StartCom---the widely trusted \emph{Certinomis - Root CA} cross-signed the \emph{StartCom EV SSL ICA} (see Figure~\ref{fig:wosign-timeline}).
The \emph{StartCom EV SSL ICA} intermediate was issued a few days before by \emph{StartCom Certification Authority G3} and thus was affected by the \emph{not before} rules set up by Google and Mozilla, rendering it unable to issue trusted certificates.
However, the new Certinomis cross-sign bypassed the \emph{not before} rules as it established a trust path to a root that was not operated by WoSign or StartCom, enabling StartCom to issue valid certificates despite its ban.
After discovery, Mozilla and Google revoked the cross-sign. Mozilla added it to OneCRL in September 2017 \cite{bugzilla_1552374,mozilla_certinomis_incidents}.
One month later, Certinomis also added the certificate to its CRL. In the end, this incident added to a list of issues that led to the distrust of Certinomis in mid 2019.
Overall, the cross-sign provided undesired valid trust paths for about \num{6} months.
In our dataset, this affected \num{11} certificates. We saw a subset of them in a small amount of connections during this \num{6} month period.
This result highlights the complexity of revocation when cross-signing is involved.
A cross-signing CA not only has to keep track of the cross-sign (as for any issued intermediate), but must also carefully examine and track actions applied to the cross-signed certificate.

The other cross-signs of WoSign/StartCom did not result in any undesired trust paths. We briefly discuss them to further stress the complexity of CA revocations given cross-signs.
WoSign received several cross-signs before its ban.
\label{special:wosign_crosssigned_by_certplus}
Keynectic's \emph{Certplus Class 1 Primary CA}, which was part of the Microsoft root store, cross-signed three WoSign CA certificates. %
These cross-signs were never revoked.
This only did not cause any undesired trust paths because Microsoft set up a \emph{not before November 16, 2016} rule for the Certplus root in September 2017 and disabled the Certplus root in May 2018.
Mozilla and Google, however, could have been more thorough when revoking WoSign and StartCom.
WoSign's roots \emph{Certification Authority of WoSign} and \emph{Certification Authority of WoSign G2} set up four (internal) intermediate \xscs{}: \emph{WoSign Class 3 OV Server CA G2}, \emph{WoSign Class 4 EV Server CA G2}, \emph{WoSign Class 4 EV Pro Server CA G2}, and \emph{WoSign Class 3 OV Pro Server CA G2}.
Despite these cross-sign's expiry not before November 2029, none of them was revoked in OneCRL or CRLSet when WoSign was distrusted.
While our data shows that this was not necessary in this case as corresponding roots are revoked, an explicit revocation can prevent undesirable trust paths---e.g. if there was another unknown cross-sign, or if WoSign was able to obtain new cross-signs for its intermediates.

Comodo explicitly revoked its cross-signs when Mozilla started to distrust WoSign in 2016.
Specifically, cross-signs of the \emph{Certification Authority of WoSign} by Comodo's \emph{UTN-USERFirst-Object} and \emph{UTN-DataCorp SGC} were revoked.
Similarly, \emph{Unizeto Certum CA} revoked its cross-sign of \emph{Certificate Authority of WoSign G2} in 2016.

\label{special:early_wosign_startcom_xs_mainpaper}
We also find that StartCom cross-signed WoSign certificates years before being acquired by them, going back to at least 2006 (details in Appendix~\ref{special:early_wosign_startcom_xs_appendix}).
This shows a close business relationship between them years before the (not timely reported) acquisition of StartCom by WoSign in 2015.
This cross-signing behavior also continued after the acquisition.
In this regard, the analysis of cross-signs can provide a rich source of information for security research.

\subsubsection{Incomplete DigiNotar Revocation} %
\label{sec:crosssigns_in_the_wild:ineffective_revocations:diginotar}

After a security incident at DigiNotar in 2011, an attacker was able to obtain certificates for several domains, including Google and Mozilla's add-on domains.
As a result, Mozilla distrusted all DigiNotar root certificates \cite{mozilla_diginotar_removal}.
Our data shows that \emph{DigiNotar Root CA} was cross-signed by \emph{Entrust.net Secure Server CA} in 2007, creating an intermediate which was valid up to August 2013.
Both, root and cross-sign, were revoked by Microsoft, Google, and Mozilla in 2011:
Microsoft removed the root from its root store and blocked the intermediate. %
Likewise, Google added both certificates to Chrome's blacklist disabling all certificates that use the corresponding public key\footnote{Technically, Chrome's blacklist lists hashes of the certificate's \emph{SubjectPublicKeyInfo} (SPKI), i.e., the encoded public key.}, including existing and possible future cross-signs.
Mozilla, as OneCRL was not introduced before 2015 \cite{mozilla_onecrl_announcement}, implemented a special revocation mechanism to disable the root cert. %
However, applications that base their trust on the certificates in the root store of Mozilla, Google, or Microsoft, often do not support their special revocation mechanisms.
As the cross-signing root \emph{Entrust.net Secure Server CA} remained in root stores until 2015 (Mozilla, Google) or longer (Microsoft, Apple), such applications still established a valid path until expiry of the DigiNotar cross-sign in August 2013.
Hence, applications were at risk to accept certificates issued by \emph{DigiNotar Root CA} up to two years after the distrust of DigiNotar roots because of the cross-sign.

\subsubsection{Incomplete Vendor-Controlled CRLs} %
\label{sec:crosssigns_in_the_wild:ineffective_revocations:crl_inconsistencies}

Most applications do not perform certificate revocation checks---especially not for CA certificates (see Section~\ref{sec:certificate_basics:crl_methods}).
Instead, many browsers and operating systems ship their own revocation lists, which they have to keep updated.
The security of the web PKI thus also depends on how consistently cross-signed certificates are revoked.
In this section, we examine these lists and show that cross-signs are \emph{not} consistently revoked. This further highlights the complexities of cross-signs.

\label{par:crosssigns_in_the_wild:intermediate_cs:extern:crl_inconsistencies}
In 2011, the \emph{Actalis Authentication Root CA} created the intermediate \emph{Actalis Authentication CA G2}.
\label{special:91513_bootstrapping_description_in_inconsistency_section}
A cross-sign by \emph{Baltimore CyberTrust Root} enabled broad trust. %
The \emph{Actalis Authentication CA G2} was later revoked by Actalis with ``cessation'' given as the reason in November 2016. The same is true for the cross-sign by CyberTrust. 
OneCRL included these revocations, but Google lists only the Actalis intermediate in its CRLSet.
Hence, the cross-sign still works for devices using Google's root store---like Android phones. This affected $13$ certificates we saw in traffic; they were recognized as valid for two years until expiry of the cross-sign.

Another revocation inconsistency affects the internal intermediate \xsc{} \emph{GlobalSign Extended Validation CA - SHA256 - G2}. \label{special:GlobalSign_revocation_inconsistency}
The first certificate issued by \emph{GlobalSign Root CA - R2} was revoked in September 2019 due to cessation of operation, but GlobalSign did not revoke the cross-sign by \emph{GlobalSign Root CA - R3}.
OneCRL (Mozilla) and CRLSet (Google) mimic this inconsistency \cite{bugzilla_1602265}.
Lacking any explanation, we believe that the cross-sign should have been revoked, too.
In this case, we did not find any non-expired certificates affected by this.

The intermediate \xsc{} \emph{Entrust Certification Authority - L1E} shows a similar revocation inconsistency:
While \emph{Entrust Root Certification Authority} revoked the corresponding intermediates in July 2018, the cross-sign by \emph{Entrust.net Certification Authority (2048)} was not revoked before February 2019, i.e., seven months later.
Google's CRLSet even missed revoking one of the earlier CRL-revoked cross-sign. %
Although no severe security problem evolved (the intermediate was intentionally superseded and all issued certificates expired before the initial revocation), both cases show that already cross-signing within a CA can inadvertently prolong the validity of certificates, let alone cross-signing across CAs.
\label{special:430454_565445_17833656_inconsistently_revoked}
Three intermediate \xscs{} of the US Federal PKI show inconsistent revocation states.
For all three (\emph{DoD Interoperability Root CA 2}, \emph{NASA Operational CA}, and \emph{DHS CA4}), one intermediate was revoked via CRL, marking them as superseded.
However, corresponding cross-sings were not revoked; without an obvious explanation.
Vendor-controlled CRLs did not even add the revoked certificates.
A total of \num{10}, \num{5725} and \num{4} certificates were affected, respectively.

Finally, special cross-signing-related requests of CAs can further complicate the maintenance of vendor-controlled CRLs.
Specifically, the root \xsc{} \emph{Belgium Root CA2} was cross-signed by \emph{GlobalSign Root CA} and \emph{CyberTrust Global Root}.
While the former cross-sign expired in 2014, Belgium requested an inclusion of the CyberTrust-issued cross-sign in OneCRL in October 2017 because the intermediate is no longer used to issue TLS certificates~\cite{bugzilla_1404501}.
This request created a very narrow-band revocation as it affects only applications that use OneCRL, i.e., mostly Firefox:
First, neither Google nor Microsoft block the cross-sign in their vendor-controlled CRLs.
Second, the root is still included in Apple's root store today (and was never included in any other) until its expiry in 2021.
Third, the cross-sign is not included in the CA's CRL, probably as CRLs lack a mechanism to revoke certificates for specific key usages (notably, the only allowed key usage for the root are \emph{Certificate Sign} and \emph{CRL Sign}).
On the positive side, we can not provide evidence for a real world relevance of this inconsistency: All observed issued certificates already expired before the (incomplete) revocation.

\textit{
\textbf{Takeaway:}
Cross-Signing resulted in incomplete revocations or even added new valid trust paths for already revoked certificates.
Thus, coping with cross-signing consequences adds a further burden to the important but fragile revocation mechanisms in PKIs.
Consequently, we propose adapted mechanisms that ease the revocation in face of cross-signing.
Furthermore, CAs should be required to publicly disclose and explain inconsistencies even if they serve a benign purpose. %
}

\subsection{PKI Barrier Breaches} %
\label{sec:crosssigns_in_the_wild:pki_barrier_breaches}
Next, we show how cross-signing can enable a valid trust path between otherwise isolated PKIs---a potentially undesired effect.
We discuss how cross-signs intentionally or inadvertently cause breaches of such PKI boundaries, e.g., by making state-controlled PKI systems trusted by the Web PKI.
We show this effect on two examples of the Federal PKI and the Swiss Government PKI.

\subsubsection{Undesired Global Trust in the FPKI} %
\label{sec:crosssigns_in_the_wild:pki_barrier_breaches:fpki}
\label{case:crosssigns_in_the_wild:fpki_breack}
\label{par:crosssigns_in_the_wild:root_cs:internal_root_xscs:fpki}

The Federal PKI (FPKI) provides PKI functionality for US government services.
While the FPKI was part of the Apple and Microsoft root programs till 2018, it never was accepted by Mozilla~\cite{mozilla_symantec_incidents,bugzilla_478418}.
However, we find several cross-signs of FPKI CAs by widely trusted Web-PKI CAs.
As the FPKI extensively uses internal cross-signs among its CA certificates, this trust in single CA certificates expanded to a large part of the FPKI.
As a result, many applications that use the Mozilla root store unknowingly trusted certificates issued by the FPKI, even though it did never fulfill the necessary root store policies~\cite{mozilla_symantec_incidents,bugzilla_478418}.
As the FPKI applied for trust by Mozilla since 2009~\cite{bugzilla_478418} (aborted in 2018), the cross-signs could be seen as attempt by the FPKI to \emph{bootstrap} (cf. Section~\ref{sec:crosssigns_in_the_wild:bootstrapping_help}) trust in Mozilla's rootstore in parallel.
However, at the time of cross-signing, the public discussion on the FKPI's application for trust already called out concerns regarding the eligibility of affected certificates~\cite{bugzilla_478418}.
These concerns should also have prevented the cross-signs. %
This is especially worrisome since
state-controlled PKI systems can suffer from potential political influence~\cite{soghoian20012}.
We start our analysis with the intermediate \xsc{} \emph{Federal Bridge CA 2013}.
This intermediate was issued in 2013 by the \emph{Federal Bridge CA} and two years later cross-signed by the \emph{Federal Common Policy CA}.
More importantly, also in 2015, it was cross-signed by widely trusted CA certificates of IdenTrust and VeriSign:
First, it was cross-signed by the intermediate \emph{IdenTrust ACES CA 1} which roots back to \emph{DST ACES CA X6}.
Second, VeriSign cross-signed the FPKI CA with its \emph{VeriSign Class 3 SSP Intermediate CA - G2}, an intermediate issued by VeriSigns internal \xsc{} \emph{VeriSign Universal Root}.
These cross-signs provided the FPKI with broad trust coverage until the cross-signs expired or were revoked.
VeriSigns cross-sign expired in July 2016, shortly before the issuing \emph{VeriSign Class 3 SSP Intermediate CA - G2} was revoked by CA and vendor-specific CRLs in 2017.
IdenTrust revoked its cross-sign in February 2016, but OneCRL did not inherit the CRL entry before November 2017.
Until revocation in 2015, a further trust path was provided via a cross-sign by the \emph{Federal Bridge CA} which was itself issued by the already aforementioned \emph{DST ACES CA X6}~\cite{bugzilla_478418}.
The thus trusted \emph{Federal Bridge CA} and \emph{Federal Bridge CA 2013} distributed the trust further by cross-signing the \emph{FPKI Federal Common Policy} root.
Between 2008 and 2011, a further trust path to \emph{DST ACES CA X6} for this FPKI certificate was established by a cross-sign from the root \xsc{} \emph{FBCA Common Policy}.
\emph{FBCA Common Policy} offers this path to \emph{DST ACES CA X6} via a cross-sign by \emph{FBCA Entrust}; additionally it is included in recent Apple root stores and trusted by Microsoft (between 2012 and 2017).
These trust paths are also available for \emph{Federal Bridge CA 2013} which was (mutually) cross-signed by \emph{FPKI Federal Common Policy}.
Further cross-signs by \emph{Federal Bridge CA 2016} and \emph{U.S Department of State AD Root} do not provide new trust paths, however, as these have in return been signed by \emph{FPKI Federal Common Policy}, too, the broad trust also expands to these intermediates.

Further cross-signs by \emph{FPKI Federal Common Policy} expand the IdenTrust and VeriSign trust path deeper in the FPKI:
It cross-signed the intermediate \emph{SHA-1 Federal Root CA}; as likewise done by \emph{FBCA Common Policy} (providing an alternative path to \emph{DST ACES CA X6}) and, again, \emph{VeriSign Class 3 SSP Intermediate CA - G2}.
\emph{SHA-1 Federal Root CA} further signed the \emph{DoD Interoperability Root CA 1} %
which repeatedly cross-signed the \xsc{} \emph{DoD Root CA2}. %
However, in contrast to the former \xscs{}, \emph{name constraint} extensions limit these cross-signs to issue for the U.S. Government only. %

\subsubsection{Cross-Signs of the Swiss Government} %
\label{sec:crosssigns_in_the_wild:pki_barrier_breaches:swiss}

In 2016, the Swiss Government created the \emph{Swiss Government Public Trust Standard CA 02} intermediate which was cross-signed by \emph{QuoVadis Enterprise Trust CA 2 G3} in 2017.
Positively, QuoVadis used \emph{X509v3 Name Constraints} to white-list domains for which these cross-signs are allowed to issue certificates.
However, QuoVadis did not set the \emph{critical} flag for this extension, i.e., implementations are allowed to ignore it.
Consequently, the cross-signs could yield undesired trust paths for software that does not implement X509v3 Name Constraints.
This especially could affect applications that derive trusted roots from root stores of Mozilla, Google, or Apple as these do not establish a valid path for the original intermediate (contrarily to Microsoft's root store).
QuoVadis used its CRL to revoke the cross-signs in mid 2019, but only CRLSet adopted this revocation. %
The Swiss Government did not misuse this opportunity: All \num{1039} certificates observed in our measurements are part of the white-listed domains.

Contrarily, \emph{Baltimore CyberTrust Root} did not set up name constraints when it cross-signed \emph{Swiss Government SSL CA 01} (an intermediate issued in 2014 by \emph{Swiss Government Root CA II} which is included in Microsoft's root store since 2016).
Until its expiry in 2017, this cross-sign allowed the Swiss CA to issue certificates with valid trust paths to all major root stores.
In total, \num{756} certificates validate under these circumstances; \num{9} of them are not part of the white-listed domains, but still part of the Swiss top-level domain.

We describe three more cross-signs of state-controlled CAs when discussing DigiCert in the light of ownership changes (Section~\ref{special:digicert:inherited_state_controlled_xs}).

\textit{
\textbf{Takeaway:}
Due to the extensive cross-signing in the FPKI, only few trust anchors to the Web-PKI added many new trust paths.
This highlights the need for mechanisms that provide CAs with better insight on the effect on trust paths before they cross-sign a certificate.
Furthermore, enforcing short validity periods for intermediates could limit the impact of unexpected trust paths.
}

\subsection{The Good: Bootstrapping of new CAs} %
\label{sec:crosssigns_in_the_wild:bootstrapping_help}

In contrast to the previous cases which show security problems of cross-signing, we now focus on the benefits of cross-signing.
Especially for new CAs, inclusion into root stores is a lengthy process.
Obtaining a cross-sign from a broadly trusted root or intermediate enables a CA to start its business while pursuing the process to include certificates into root stores.
In our dataset, we can identify such bootstrapping help for Let's Encrypt, the China Internet Network Information Center (CNNIC), and GoDaddy. %
Furthermore, CyberTrust bootstrapped trust in \emph{Actalis Authentication CA G2} when Actalis' root had not yet been trusted (cf. Section~\ref{special:91513_bootstrapping_description_in_inconsistency_section}).
We provide a full list of involved CAs, e.g., COMODO, Digicert, WoSign, Globalsign, AffirmTrust, and government CAs, in Appendix~\ref{app:cross_signs:bootstrapping_further_examples}.

\subsubsection{Bootstrapping Let's Encrypt} %
\label{sec:crosssigns_in_the_wild:bootstrapping_help:lets_encrypt}

Let's Encrypt launched as nonprofit Certificate Authority in 2015.
It significantly increased the amount of secured Internet communication by automating the certificate issuing process and providing certificates for free.

However, to provide this service early on, it had to depend on a cross-sign by a widely trusted root for more than 5 years.
Specifically, Let's Encrypt will first start to default to its own ISRG Root starting September 29, 2020 \cite{letsencrypt-crosssign-transition}.
Before, IdenTrust helped to bootstrap trust for Let's Encrypt.
To this end, its \emph{DST Root CA X3} cross-signed four intermediates originally issued by the \emph{ISRG Root X1}, a root of the Internet Security Research Group (ISRG) which manages Let's Encrypt.
As \emph{ISRG Root X1} was initially not included in root stores, IdenTrust was the sole trust anchor for Let's Encrypt certificates enabling a fast ramp up of the service \cite{letsencrypt-crosssign}.

Looking at the details, Let's Encrypt uses four intermediate \xscs{}.
\emph{Let's Encrypt Authority X1} and \emph{X2} are no more actively used \cite{letsencrypt-crosssign} (but not revoked and still valid until the end of 2020).
In contrast to newer cross-signs, IdenTrust prevents the X1 and X2 cross-signs from issuing certificates for the top-level domain \emph{.mil}.
The more recent \emph{Let's Encrypt Authority X3} is currently used by Let's Encrypt to issue (leaf) certificates and remains valid until 2021-10 (original) and 2021-03 (cross-sign).
Finally, \emph{Let's Encrypt Authority X4} serves as backup, but will expire at the same time as X3.

Even after the switch to the own ISRG roots, the cross-signs will be beneficial:
Legacy clients that do not include the ISRG roots in their root store can fall back to the IdenTrust trust path~\cite{letsencrypt-crosssign}.

\subsubsection{Entrust Helped CNNIC} %
\label{sec:crosssigns_in_the_wild:bootstrapping_help:cnnic}

Similar to Let's Encrypt, the Chinese CNNIC obtained bootstrapping help from an established CA;
\emph{Entrust.net Secure Server Certification Authority} issued a \emph{CNNIC SSL} intermediate in 2007.
Shortly after the inclusion of \emph{CNNIC ROOT} into root stores (e.g. 2009 for Mozilla \cite{bugzilla_527759}), the \emph{CNNIC ROOT} cross-signed this intermediate, creating an alternate trust path.

\subsubsection{GoDaddy -- Internal Bootstrapping via Subsidiary} %
\label{sec:crosssigns_in_the_wild:bootstrapping_help:godaddy}

Similarly, GoDaddy used cross-signing to bootstrap trust into its CA certificates \cite{bugzilla_403437} when it entered the certificate business in 2004.
In contrast to the previous cases, however, it used an \emph{internal} cross-sign.
Particularly, GoDaddy started off with the root certificates \emph{GoDaddy Class 2 CA} and \emph{Starfield Class 2 CA} and cross-signed them with ValiCert's root \emph{ValiCert Class 2 Policy Validation CA} (created 1999).
Interestingly, GoDaddy acquired ValiCert just the year before, probably already preparing for its new business.
Until its removal from root stores around 2014, the ValiCert root thus bootstrapped trust to Mozilla, Android, and Apple for the new certificates.
GoDaddy later bootstrapped a root for Amazon: When Amazon created its \emph{Amazon Root CA 1} in 2015, an immediate cross-sign by \emph{Starfield Services Root CA - G2} established trust right away whereas it took years for Amazon's root to arrive in root stores. E.g., Mozilla and Google included it in 2017 and Apple even later in 2018.

\textit{
\textbf{Takeaway:}
Cross-Signing enables new CAs to already start their business during the process of including their roots into root stores.
Without such a cross-sign, the long periods for inclusion and sufficient propagation of root stores of several years could be prohibitive for new companies in this business.
}

\subsection{Expanding Trust and Alternative Paths} %
\label{sec:crosssigns_in_the_wild:trust_coverage_and_fallback_trust}
Any cross-sign either (i) \emph{expands the trust to more root stores}, i.e., provides trust paths to not yet covered root stores,
(ii) \emph{extends the validity period} in covered root stores,
or (iii) \emph{adds alternative paths} to root stores already covered by other certificates in the \xsc{}.
A \xsc{} often causes several of (i)--(iii); Table~\ref{tab:crosssigns_in_the_wild:quantitative_overview} only counts them to their primary case (i $>$ ii $>$ iii).
We briefly describe some \xscs{} in the following and refer to Appendix~\ref{app:crosssigns_in_the_wild:trust_coverage_and_fallback_trust} for further examples.

\subsubsection{Expanding Trust} %
\label{sec:crosssigns_in_the_wild:trust_coverage_increase}

Some root certificates are not included in all major root stores.
We find that CAs broadly use cross-signs to close these holes in the root-store coverage, enhancing their trust.
E.g., a cross-sign %
provides \emph{Entrust Root Certification Authority - G2} with trust paths for Mozilla. %
Similarly, USERTrust cross-signed \emph{AddTrust Qualified CA Root} and \emph{AddTrust Class 1 CA Root} to provide trust for Microsoft and the grid PKI.
On the downside, as aforementioned, these additional trust paths complicate certificate revocations.
Bootstrapping (cf. Section~\ref{sec:crosssigns_in_the_wild:bootstrapping_help})
is a special case of this trust-expanding cross-signing.
The count in Table~\ref{tab:crosssigns_in_the_wild:quantitative_overview} excludes bootstrapping cases.
Furthermore, we distinguish cases that extend only the validity period.
E.g., \emph{GlobalSign Domain Validation CA - SHA256 - G2} originally used trust paths via \emph{GlobalSign Root CA - R3}, while a cross-sign by \emph{GlobalSign Root CA} extends the trust by several years.

\subsubsection{Alternative Paths} %
\label{sec:crosssigns_in_the_wild:fallback_trust}

Often, multiple certificates of a \xsc{} provide valid paths to the same root store.
This approach provides fall-back trust: it proactively maintains alternative paths to deal with unexpected revocations or removals.
E.g., the intermediate \xscs{} for \emph{Servision Inc.}, \emph{XiPS}, and \emph{KAGOYA JAPAN Inc.} -- originally issued by GoDaddy's \emph{ValiCert Class 1 Policy Validation Authority} -- were cross-signed by SECOM's \emph{Security Communication RootCA1} in 2012.
The cross-signs provided readily usable fall-back paths when the originally issuing ValiCert certificate was removed from Mozilla and Google root stores due to its 1024 bit RSA key~\cite{bugzilla_936304,bugzilla_881553}.

Note that most cross-signs establish alternative paths as issuing CA certificates are often trusted in many root stores.
The count in Table~\ref{tab:crosssigns_in_the_wild:quantitative_overview} only lists \xscs{} whose sole outcome are alternative paths, i.e., it includes only \xscs{} that do \emph{not} expand the trust.

\textit{
\textbf{Takeaway:}
Cross-Signing enables large root store coverage if issuing CAs span only a subset of root stores.
Likewise, cross-signing can provide alternative trust paths to proactively deal with CA revocations and removals.
Both ensures a non-disruptive user experience.
However, multiple trust paths for a certificate can also lead to incomplete revocations and thus challenge the security of PKIs (cf. Section~\ref{sec:crosssigns_in_the_wild:ineffective_revocations}).
Using cross-signing for these purposes thus necessitates better mechanisms to mitigate these security problems.
We discuss possible solutions such as better logging and limited lifetimes for \xscs{} in Section~\ref{sec:discussion}.
}

\subsection{Cross-Signing Eases the Transition to New Cryptographic Algorithms} %
\label{sec:crosssigns_in_the_wild:changing_algorithms}

Security guidelines by entities like the CAB Forum motivate CAs to support advancements in cryptography early on, e.g., new signature algorithms.
To maintain backward compatibility for legacy implementations, CAs use cross-signs to establish alternative trust paths that support new algorithms.
Legacy clients that do not support the new algorithms can still use the old trust paths.
We find this happening commonly with intermediate \xscs{}.
For root certificates this approach is not required - their self-signed signatures are typically not checked by clients~\cite{bugzilla_650355}.
In these cases, the CA can just issue a new intermediate that uses the new algorithms.
To give some examples, the \emph{Virginia Tech Global Qualified Server CA} intermediate was issued with a SHA1 signature by \emph{Trusted Root CA G2} (GlobalSign) in 2012.
In December 2014, i.e., approaching the deadline for deprecating issuance based on SHA1 \cite{cab_ballot118}, GlobalSign cross-signed the intermediate from the \emph{Trusted Root CA SHA256 G2}. The SHA1 intermediate was revoked in January 2017.

Other intermediates switched to SHA2 long before the official deprecation:
The US Government (\emph{Common Policy}) issued a SHA1 intermediate \emph{Betrusted Production SSP CA A1} in 2008 and created a SHA256 cross-sign in 2010 using \emph{Federal Common Policy CA}.
The SHA1 intermediate was revoked in 2011.
Actalis obtained a SHA1 cross-sign from \emph{Baltimore CyberTrust Root} for its SHA256 intermediate \emph{Actalis Authentication Root CA} at time of original issuance.
Likewise, the Japanese Government issued a cross-sign for \emph{ApplicationCA2 Sub} %
to provide parallel support for SHA1 and SHA256 on creation of the intermediate in 2013.

Keynectis cross-signed its intermediate \emph{KEYNECTIS Extended Validation CA} with roots of its subsidiaries Certplus (\emph{Class 2 Primary CA}) and OpenTrust (\emph{OpenTrust CA for AATL G1}).
The cross-signs offer SHA1, SHA256, SHA512 and ECDSA-with-SHA384 signatures.

\textit{
\textbf{Takeaway:}
Cross-signing can establish new trust paths that fully support new cryptographic algorithms.
This enables state-of-the-art clients to achieve better security but at the same time maintains backward compatibility for legacy clients which validate the same certificate with the help of older paths.
We expect CAs to put similar backward compatibility procedures in place once the use of a successor for SHA2, like SHA3, becomes popular.
However, we posit that backward compatibility should be kept for a limited time only---using short certificate validity periods---to push application developers to support state-of-the-art cryptography.
}

\subsection{Effect of Ownership-Changes on \xscs{}} %
\label{sec:crosssigns_in_the_wild:ownership_changes}

In this section, we analyze the effect of CA acquisitions on existing cross-signs.
Cross-signs, especially across organizations, add contractual obligations---and necessitate trust granting CAs to check the actions of CAs they cross-signed.
Thus, we are interested if new owners revoke cross-signs that were created by acquired CAs.
Similarly, we are interested if cross-signs are revoked once the owner of a cross-signed CA changes.

\subsubsection{Cross-Signs Outlive Ownership-Changes} %
\label{sec:crosssigns_in_the_wild:ownership_changes:outliving}

The \emph{Network Solutions Certificate Authority} root has been repeatedly cross-signed by members of Comodo's trust network. This process spanned several ownership changes.
The original root was created 2006. At this time, Network Solutions had been owned by Pivotal Equity for three years.
In the same year, Comodo's \emph{UTN-USERFirst-Hardware} cross-signed the root twice.
After Network Solutions was sold to General Atlantic, Comodo's \emph{AddTrust External CA Root} cross-signed the certificate in 2010.
Furthermore, we find a cross-sign by AddTrust: this intermediate has been valid since 2000---when Network Solutions was owned by VeriSign---but it likely was backdated (cf. Section~\ref{sec:crosssigns_in_the_world:misc:addtrust_backdating}).
All these cross-sings outlived the acquisition of Network Solutions by web.com in 2011 and remain valid until 2020, thus spanning at least two ownership changes.

We also find cases in which the CAs that issue cross-sign certificates changed owners.
\label{special:digicert_xs_by_entrustDatacard}
When Digicert created the \emph{DigiCert High Assurance EV Root CA} in 2006, it was cross-signed by \emph{Entrust.net Secure Server Certificate Authority} and \emph{Entrust.net Certificate Authority}.
\num{3} years later, Thoma Bravo acquired Entrust and sold it to Datacard in 2013, which rebranded it to Entrust Datacard.
Despite these ownership changes, the cross-signs of Digicert's root remained valid (until 2014 and 2015), making the new owners responsible for trust paths of certificates issued by DigiCert\footnote{Note that Thoma Bravo was not in control of Digicert and Entrust at the same time. Thoma Bravo acquired Digicert in 2015, two years after selling Entrust to Datacard.}.
This raises the question if the new owners were aware of the cross-signs - and decided to keep them, or if they simply were not aware of their existence.

\subsubsection{DigiCert -- Internal Islands and External Legacy Cross-Signing} %
\label{sec:crosssigns_in_the_wild:ownership_changes:digicert}

In the DigiCert group, most \xscs{} were created before DigiCert acquired the corresponding CAs.
Thus, we predominantly find internal cross-signs within each subsidiary and only few cross-signs across DigiCert subsidiaries.
Most cross-signs originate from times of VeriSign, Verizon, and QuoVadis.
DigiCert only occasionally used the acquired certificates to cross-sign its own \emph{DigiCert} roots.

When acquiring Verizon in 2015, DigiCert also became responsible for external \xscs{}.
First, Verizon cross-signed root certificates of WellsFargo in 2013 and 2015.
All corresponding intermediates were revoked by the CA's CRL and Mozilla's OneCRL in 2017, when the roots were removed from all root stores (after request by WellsFargo \cite{bugzilla_1332059}), except for Apple.
Similarly, Verizon cross-signed \emph{Certipost E-Trust Primary Normalized CA} providing this former only Microsoft-trusted root a broad trust coverage.

\label{special:digicert:inherited_state_controlled_xs}
Verizon also cross-signed state-controlled CAs which is potentially problematic (cf. Section~\ref{sec:crosssigns_in_the_wild:pki_barrier_breaches}).
In 2010 and 2013, it cross-signed the \emph{Swiss Government} root, increasing its trust beyond Apple and Microsoft. %
Similarly, it cross-signed the \emph{Belgium Root CA2}
and Portugal's \emph{SCEE ECRaizEstado}--some of them were later revoked in 2018 due to a series of misissuances \cite{bugzilla_1432608}.
The acquisition of Verizon made Digicert responsible for these cross-signs of state-controlled CAs.
Thus, cross-signs did not only provide state-controlled CAs with large trust coverage, but these cross-signs also faced ownership changes which increase the risk for unnoticed problems.

We describe further \xscs{} with ownership changes and provide details on Digicert's cross-signs in Appendix~\ref{app:cross_signs:further_examples_with_ownership_change}.

\textit{
\textbf{Takeaway:}
Considering the frequent cross-signs across CAs, potential new owners must---before acquiring a CA---review existing issued and received cross-signs and corresponding obligations.
Similarly, cross-signing CAs must be informed when a cross-signed CA changes its owner.
Both requires an easily accessible and verifiable store of \xscs{} as we suggest in Section~\ref{sec:discussion}.
}

\subsection{The Ugly: Potentially Problematic Practices and Missing Transparency} %
\label{sec:crosssigns_in_the_wild:misc}

In this section, we highlight practices of CAs that---while not explicitly forbidden---are frowned upon by root store maintainers~\cite{mozilla_problematic_practices} or make it hard to assess the legitimacy of existing trust.

\subsubsection{Backdating of Cross-signs} %
\label{sec:crosssigns_in_the_world:misc:addtrust_backdating}

Comodo's \emph{AddTrust External CA Root} backdates several cross-signs, i.e., it sets the \emph{not before} field to a date several years before the actual issuance.
We used \href{https://crt.sh}{crt.sh} to verify that the early \emph{not before} dates are not caused by re-issued root certificates.
Backdating is explicitly forbidden if it bypasses a root store policy that relies on information in the \emph{not before} field---like the deprecation of SHA-1 signatures for new certificates \cite{cab_ballot118}.
Mozilla lists backdating as problematic practice~\cite{mozilla_problematic_practices} as backdating can interfere with future policies (cf. not before rules, Section~\ref{special:not_before_rule}).

The \emph{USERTrust ECC Certification Authority} root (valid since 2010) was cross-signed by \emph{AddTrust External CA Root} in 2013~\cite{bugzilla_1418148} using a \emph{not before} date of 2000.
Other cross-signs of the root use more accurate times, e.g., a cross-sign by \emph{AAA Certificate Services} in 2019.

We find further cross-signs by AddTrust which we believe to be backdated as their \emph{not before} date precedes that of the cross-signed root by several years.
We lack clear evidence, but signs for a reoccurring backdating behavior of the CA are present:
All presumably backdated cross-signs are issued by \emph{AddTrust External CA Root} using \emph{not before} dates from 2000 or 2010.
Specifically, the cross-signs of \emph{COMODO Certification Authority}, \emph{COMODO ECC Certification Authority} %
\emph{COMODO RSA Certification Authority}, and \emph{USERTrust RSA Certification Authority} %
list \emph{not before} dates from 2000 whereas the corresponding roots list dates from 2006, 2008 and 2010.
Cross-signs of the \emph{Network Solutions Certificate Authority}, a root created in 2006, list dates from 2000 and 2010.

Finally, we find StartCom to set \emph{not before} to 2006 (corresponding to the original root) when cross-signing \emph{StartCom Certification Authority} whereas the cross-signing \emph{StartCom Certificate Authority G2} is valid only since 2010.
This case is only a problematic practice and no strict violation of the policy.
However, it is noteworthy that StartCom already backdated certificates long before the incidents that led to the removal of trust in WoSign and StartCom. That decision was partly based on their violation of policies by backdating certificates to circumvent SHA1 deprecation (cf. Section~\ref{sec:crosssigns_in_the_wild:ineffective_revocations}).
\subsubsection{Missing Insight for Partly Disabled Comodo Root \xscs{}} %
\label{sec:crosssigns_in_the_wild:misc:comodo_partly_revoked}

Comodo requested the exclusion of some root certificates from root stores, but kept their cross-signs valid.
Specifically, the root certificates of \emph{USERTrust UTN-USERFirst-Hardware} and \emph{AddTrust Class 1 CA Root} were removed from the root stores of Firefox and Android in 2017 as they are no longer in use~\cite{bugzilla_1378334}.
However, corresponding cross-signs by \emph{UTN-USERFirst-Client Authentication and Email} (also trusted by Firefox and Android) have not been revoked as they are still in use.
Especially the \emph{USERTrust UTN-USERFirst-Hardware} cross-sign is still heavily used after the withdrawal of its root: We find valid trust paths for \num{4244104} leaf and \num{276} CA certificates until 2020. %
We note that, from everything we can tell, this is---in this case---desired behavior and not malicious.
It shows, however, that it is hard to tell for an outside observer if cross-signs just have been forgotten - or are purposefully kept active.

The removal (and eventual revocation) of \emph{USERTrust UTN - DATACorp SGC} is a counterexample.
It was removed from Firefox and Android root stores in 2015 due to a planned removal of public trust by the owner~\cite{bugzilla_1208461,bugzilla_1233408}.
After additional request by the owner, the intermediates were revoked by adding them to Mozilla's OneCRL~\cite{bugzilla_1233408}.
The need for an additional request shows the complexity of revocations in face of cross-signs. In our opinion, the original removal request should already have raised questions on the revocation of the cross-signs. Instead, this required an explicit additional request by the owner---which only happened about 3 months later.

\textit{
\textbf{Takeaway:}
The backdating of certificates might be tempting, e.g., to match the validity periods of the original certificate.
However, backdating is a problematic behavior that hides the issuance date of a certificate without providing benefits.
On top of this, cross-signing adds a new problem: The legitimacy of an incomplete revocation is hard to assess as corresponding information is typically unavailable.
To achieve transparency, evaluating cross-signs and requesting information about them should be an integral part of revocation processing.
}

\textit{
\textbf{\xscs{} in the Wild -- Summary:}
While cross-signing complicates the PKI, CAs also use it to adopt more secure cryptography while maintaining compatibility with legacy software.
In the following, we propose several changes with the goal of establishing secure cross-signing practices and making cross-signs comprehensible. %
}

\section{Discussion and Recommendations} %
\label{sec:discussion}

CAs depend on cross-signing.
For example, the fast tremendous success of Let's Encrypt would not have been possible without bootstrapping the CA through cross-signing, and also GoDaddy started with a cross-sign.
Cross-signing also enables CAs to issue certificates with new, more secure cryptographic algorithms while maintaining compatibility with legacy applications---as it happened with the SHA1 to SHA256 transition.
Cross-signing can be used to increase root-store coverage by establishing paths to different root stores---even in face of differing root store policies and resulting diverging root store setups.
Cross-signing can pro-actively set up alternative trust paths based on other CA certificates such that end-user certificates still validate.
Thus, cross-signing is a beneficial mechanism for the PKI ecosystem.

Yet, cross-signing complicates certificate handling and introduces new problems, some concerning the management of cross-signs.
For an observer it is hard to tell if cross-signs intentionally or unknowingly outlive CA owner-ship changes (see Section~\ref{sec:crosssigns_in_the_wild:ownership_changes:outliving}).
Problematic practices like backdating of certificates~\cite{mozilla_problematic_practices} further complicate the assessment of existing cross-signs.
The by most important problem are undesired trust-paths caused by cross-signs.
While revocation of CA certificates already is a complicated process, necessitating software updates and/or updates of vendor-specific CRLs, cross-signs make this process even more complex---and led to numerous incomplete revocations as shown in Section~\ref{sec:crosssigns_in_the_wild:ineffective_revocations}.
New cross-signs need careful examination to make sure that they do not accidentally extend the trust of a CA, or span trust across PKIs.
Cross-sign revocations are a hard to solve problem for application software.
For example, all major browsers and Microsoft use their own systems to propagate CA revocation information.
This revocation information is, however, not easily accessible for many applications.
A typical Unix program that uses OpenSSL or NSS has no easy way to get revocation information. It will typically blindly trust all CA certificates that it encounters---no matter if it was revoked.
The same is true for mobile applications on iOS and OSX. Even on Microsoft Windows one will only get access to some revocation information---and only when using the operating system's APIs. Many applications, however, chose to use OpenSSL even on Windows---e.g., through using the popular libcurl~\cite{curlschannel}. Programming languages like Go and Java also have their own validation logic and will thus not get revocation information~\cite{javasecuresocketextensions}.

Thus, cross-signing puts the security and privacy of users at risk:
It can undermine the security of one of the most critical infrastructures in a digitally interconnected world, which---for better or worse---is the current trust backbone for nearly all secure communication on the Internet.

In the following, we suggest best practices preserving the benefits of cross-signing while limiting future security problems (Sections~\ref{sec:discussion:limit_impact_of_issues} and \ref{sec:discussion:better_path_validation_practices}) and ensuring \emph{awareness} of CAs as well as giving more \emph{transparency} to the user (Sections~\ref{sec:discussion:awareness_rules_first}--\ref{sec:discussion:awareness_rules_last}).
Automated checks can hint at violations of these practices, but---like for many rules set up by root store maintainers---manual inspection is often necessary. %

\subsection{Limit Problems using Short Validities} %
\label{sec:discussion:limit_impact_of_issues}
\label{sec:discussion:short_validity_periods}

Typically, cross-signs have validity periods ranging from several years up to several decades.
As a result, security issues often remain problematic for years---especially for applications that do not use revocation information.

To limit the impact of security incidents related to cross-signing, we suggest CAs to limit the maximum validity period of cross-signs.
The benefits of shorter validity periods have been discussed, e.g., in \cite{Topalovic2012_towards,Chuat2020_sok}.
Applications that lack access to revocation information will benefit greatly from this change---there is a much smaller window of opportunity for exploitation after a CA certificate should have been revoked.
Continuous renewals of CA certificates will also limit the validity of cross-sign to the term during which a business relationship between two entities is ongoing---such cross-signs cannot just be forgotten after an ownership change.
Short validity periods for cross-signs should not pose an insurmountable problem for CAs: Increasing automation of certificate issuance and renewal---especially advanced by Let's Encrypt in the recent years---already enables the deployment and update of short-living leaf certificates on end-user systems like webservers.
Let's Encrypt already defaults to a 3-month validity period for leaf certificates.
Our proposal also matches current industry trends: Mozilla, Google, and Apple already announced to limit the maximum validity period of leaf certificates to 398 days (1 year + grace period)~\cite{google_398limit_on_certs,mozilla_398limit_on_certs,apple_398limit_on_certs}.
To lessen the impact of problems even more, validity periods should be reduced to about 4 days \cite{Topalovic2012_towards,Chuat2020_sok} and---as underlined by our findings---also apply to CA certificates \cite{Topalovic2012_towards}. This would necessitate all CAs to adopt automation similarly to Let's Encrypt.

A more invasive and less preferable approach could be to
couple a long-living cross-sign with short-lived proofs of freshness provided by the cross-signing CA.
Whenever the certificate is used, the receiver checks if the proof of freshness is recent.
These freshness proofs can be distributed by the webserver in the TLS handshake using OCSP stapling---which can be mandated by the certificate using the OCSP Must-Staple extension~\cite{RFC6961,RFC7633}.
However, OCSP Must-Staple does not provide the necessary deployment at CAs, servers, or clients \cite{Taejoong2018_ocsp,Chuat2020_sok}.
Thus, short validity periods---which do not require specific client support \cite{Topalovic2012_towards,Chuat2020_sok}---are preferable.

\subsection{Shedding Light on Cross-Sign Motivations} %
\label{sec:discussion:shed_light_on_cross-sign_reason}
\label{sec:discussion:awareness_rules_first}

Our analysis also shows good and positive use cases for cross-signing.
However, CAs are not required to state their motivation for a particular cross-sign, although this motivation is important for assessing if a cross-sign (still) serves a benign purpose or should be revoked.
For example, a bootstrapping cross-sign can be revoked when the original root has reached the desired trust coverage.
Similarly, a cross-sign that improves root store coverage should be revoked when the original CA certificate is no longer used.
To shed light on the motivation and enable corresponding checks, we suggest requiring CAs to encode the motivation(s) for a cross-sign as a new \emph{XS extension} in the resulting certificate as follows.

\textit{Bootstrapping.} %
The XS extension should encode a reference to the bootstrapped certificate as well as the targeted root stores of the bootstrapping.
A corresponding pending request for inclusion of the certificate into the root store must exist---root store maintainers can provide a proof.
The cross-sign must not be renewed any longer when the bootstrapped certificate is trusted by the targeted stores.

\textit{Expanding Trust.} %
A cross-sign that expands the trust should encode the corresponding stores in its XS extension.
Such a certificate must only exist if no other certificate of the \xsc{} provides a valid path to the targeted root store (except for paths that are flagged as alternative, see below).
To limit the impact of coincidentally resulting alternative paths by the cross-sign, applications may ignore cross-signs that intend to expand the trust to only other root stores than the store used by the application.

\textit{Fall-Back paths.} %
While fall-back paths increase the complexity of the PKI, they can be necessary to, e.g., support legacy devices using old versions of trust-stores with old root certificates.
Fall-back cross-signs should encode the targeted root stores and the current certificate they are a fall-back for.
We posit that software that ships with up-to-date root stores that get updated regularly should ignore all fall-back certificates.
As we describe, other cross-sign use cases must prove that they do \emph{not} serve the sole purpose to create fall-back paths---via verifiable information in the XS extension.

\textit{Multiple Algorithms.} %
When providing support for new (signature) algorithms, the cross-sign must---upon issuance---establish a valid path to a root store using the desired algorithms only.
No other certificate should provide a valid path to the root store using these algorithms, except for fall-back paths (see above). %
To enable checks of this condition, the XS extension should encode the set of algorithms which are used to validate the new path; it should also encode the other CA certificates for this path.

\textit{Creation time.} %
In addition to the XS-extension, all newly issued CA certificates should include signed timestamps of several CT logs. These can serve as indicator for the time of issuance (rather than the easily manipulatable \emph{not before} field).

\subsection{Ownership Change: Report on Cross-Signs} %
\label{sec:discussion:require_action_on_ownership_changes}

As our analysis in Section~\ref{sec:crosssigns_in_the_wild:ownership_changes} shows, cross-signs typically outlive ownership changes of involved CAs.
However, the public remains unaware if this is an explicit decision.
We suggest providing transparency and raising the awareness for cross-signs when owners change.
To this end, when the owner of a cross-sign changes, the issuers of the cross-sign should be required to publicly declare if the cross-sign should remain valid.
Likewise, the new owner should publicly acknowledge that it will adhere to the obligations that arise from the possession of the cross-sign.
We envision such statements to become a mandatory part of the report on CA ownership change that is already required by root stores~\cite{mozilla_roostorepolicy}.
Incomplete statements on affected cross-signs would justify revocations and challenge the trustworthiness of the misbehaving CAs.

\subsection{Explain Revocation Inconsistencies} %
\label{sec:discussion:explain_revocation_inconsistencies}

We uncover several revocation inconsistencies for \xscs{} (cf. Sections~\ref{sec:crosssigns_in_the_wild:ineffective_revocations:crl_inconsistencies} and \ref{sec:crosssigns_in_the_wild:misc:comodo_partly_revoked}).
However, it is sometimes unclear if these inconsistencies are accidental---or desired.
To enable root store maintainers and researchers to correctly classify these cases, CAs should be required to pro-actively explain revocation inconsistencies of \xscs{}---and explain their purpose to allow for the setup of checks (cf. Section~\ref{sec:discussion:shed_light_on_cross-sign_reason}).
This information can nicely extend the existing best practice to explain the reason for a revocation in CRLs.

\subsection{Enable Easy Access to Cross-Signs} %
\label{sec:discussion:easy_access_to_cross-signs}
\label{sec:discussion:awareness_rules_last}

Easy access to cross-signing information would significantly help to ensure that cross-signs are (still) legitimate.
Certificate Transparency (CT) already aims at making certificate information available to everyone~\cite{RFC6962}.
CT provides semi-trusted append-only public log-servers with the goal to contain all currently trusted certificates in the Internet---and it is well on the way to satisfying this goal~\cite{scheitle2018_ct}.
CT is, however, not easily searchable and fragmented in lots of servers. Due to scalability problems, this fragmentation will likely increase in the future~\cite{scheitle2018_ct}.
Thus, current CT practices do not ensure a great fit of CT for the evaluation of cross-signs.

Fragmentation of certificates to different CT logs complicates the search for cross-signs since they can be distributed across different logs.
We suggest requiring CAs to report all certificates of a \xsc{} to the same CT log(s). %
This ensures that a single CT log can provide full information on the cross-signing properties of a certificate.
To account for the high impact of root and intermediate \xscs{}, CAs should be required to report all CA certificates to a set of CT logs that only log CA certificates.
These CT logs can provide fast and complete statements on the cross-signing properties of CA certificates.
Due to the limited amount of CA certificates, they should remain small---and not require a huge infrastructure commitment.

\subsection{Use Vendor-Controlled CRLs by Default} %
\label{sec:discussion:better_path_validation_practices}

In Section~\ref{sec:crosssigns_in_the_wild:ineffective_revocations}, we show the importance of revocation information for intermediate certificates.
We also highlight that, for the most part, only large webbrowsers get this revocation information---via vendor-specific CRLs.
This, however, opens a potential window of attack on all other applications which lack a sufficiently simple way to get this revocation information \cite{Oneill2017_trustbase}. %

We propose that a new standard system for the revocation of certificates should be created---which could be equivalent to either OneCRL or CRLSets. This system should be supported out-of-the-box by typical operating systems and SSL/TLS libraries (like OpenSSL). It should not necessitate any changes in applications using TLS.
TrustBase~\cite{Oneill2017_trustbase} could be a candidate.
Further, recent research shows the applicability of local revocation lists like OneCRL even for an efficient revocation of leaf certificates~\cite{larisch2017_crlite}.
Making such approaches the default would significantly improve the safety and robustness of the whole Web PKI.

\section{Related Work} %
\label{sec:related_work}

While there is a large body of work that examines different facets of the PKI~\cite{holz2011_sslLandscape, Scheitle2018_first, szalachowski2016_pki, amann2017_mission, lee2018_towards} as well as the TLS and the HTTPS ecosystem~\cite{Durumeric2015_email, Huang2014_sslcerts}, most studies do not mention or examine cross-signing.
In their 2013 Systematization of Knowledge paper, Clark and Oorschot give a thorough review of the issues of the Web PKI~\cite{clark2013_sok}.
In this paper, the authors already note the problem of revoking CA certificates. They also mention that CA certificates may not even contain the necessary information for revocation checking.
One of the first studies of the HTTPS ecosystems performed by Durumeric \etal already noted a high occurrence of cross-signing among CA certificates in 2013 \cite{Durumeric2013_analysis}.
However, their analysis is limited to an overview on the occurrence of general cross-signing and a very brief statement on effect on root store coverage without any further analysis.
Acer~\etal \cite{acer2017_wild} analyzed the causes for certificate validation errors encountered by Chrome users during web browsing.
They find that missing cross-sign certificates in a presented certificate path cause some errors and thus acknowledge the importance of cross-signing as means to root trust in widely trusted stores.
Roosa~\etal \cite{roosa2013_trust} identified cross-signing as one of the mechanisms that cause intransparency in PKI systems as CAs do not disclose cross-sign relationships when applying for inclusion in a root store.
Only since April 2018, CAs must report all intermediates that provide a path to a root in Mozilla's root store and are not constrained to specific domain subtrees \cite{mozilla_november2017communication,mozilla_roostorepolicy2_5}.
Casola~\etal \cite{Casola2005_policybased_crosscertification} enable CAs to automatically check the compatibility of their policies \emph{before} cross-signing.
We highlight problems \emph{beyond policies} and provide guidelines for a safe \emph{operation} of cross-signs.

There is a large amount of work that examines different aspects of certificate revocation.
Most studies, however, concentrate on the revocation of end-host certificates and do not detail the impact of revocation of intermediates or the interaction with cross-signing.
Liu~\etal measure certificate revocation in the Web PKI \cite{liu2015_e2e}.
While they mention the importance of revoking intermediate certificates, their study only measures leaf-certificate revocations.

\section{Conclusion} %
\label{sec:conclusion}

Our longitudinal study shows that cross-signing is a common practice in the Web PKI.
We provide a classification of possible cross-sign patterns and analyze their use in and their effect on real world communications. %
We show that cross-signs are used to bootstrap trust in new CAs---which played a significant role in the tremendous success of Let's Encrypt.
Cross-signs also lead to better trust-store coverage, giving CAs trust in stores they are not directly trusted in.
Finally, cross-signing allows for a graceful transition to newer cryptographic algorithms like done in the case of SHA-2.

However, our work also highlights the problems that cross-signing introduces to PKIs, the largest being unwanted trust paths.
Cross-signing complicates the already error-prone revocation processes of certificates. It led to numerous incomplete revocations.
We also show that introducing new cross-signs requires extensive checks to prevent accidental PKI boundary breaches. %
Cross-signing also makes it hard for observers to determine if trust paths are legitimate---or just forgotten artifacts of past contractual relationships between CAs.

Based on the gathered insights, we propose new cross-signing best practices. It is our hope to initiate and steer the discussion for new rules that preserve the beneficial potential of cross-signing but mitigate its risks.

\begin{acks}

We thank the anonymous reviewers for their valuable feedback.
This work was supported by the US National Science Foundation under grant CNS-1528156. Any opinions, findings, and conclusions or recommendations expressed in this material are those of the authors or originators, and do not necessarily reflect the views of the National Science Foundation.

\end{acks}

\newpage
\appendix

\section{Ethics of data collection}
\label{app:data_ethics}

The passive data collection effort that we perform has been cleared by the respective
responsible parties at each contributing institution before they begin contributing. Our
data-collection effort generally leverages the already existing network security monitoring
infrastructure at the contributing institutions: the data collection effort is generally run by the
network security teams of the contributing institutions on their already existing network monitoring
platforms. We just provide them with a short script (which they can examine) that performs the
TLS data extraction.

This script only extracts data that is not privacy sensitive - or anonymizes it \emph{before} the
data is sent to us. For example, it does \emph{not} include the client IP address. Instead,
the data that we are sent contains a seeded hash of the concatenation of the client and the server IP addresses.
The seed is site-specific and unknown to us.

This approach allows us to determine when the same client connects to the same server repeatedly (e.g.\ to
evaluate the effectiveness of session resumption), without enabling us to track which sites
a single client accesses.

Moreover, note that we also do not include client-certificates in our data collection; we only collect
certificates that Internet servers send to clients.

\section{Details and Further Examples} %
\label{app:cross_signs}

In the following, we detail some already presented \xscs{} and provide further examples for some cross-signing categories.
\iflongversion
\else
We describe even more examples in the extended paper version\footnote{\fix{\url{https://github.com/pki-xs-analysis/cross-signing-analysis/blob/master/README_extended_paper_version.md} \todo{ArXiv URL}}}.
\fi

\subsection{WoSign \& StartCom: Early Cross-Signing} %
\label{app:cross_signs:startcom_wosign}

\label{special:early_wosign_startcom_xs_appendix}
As briefly noted in Section~\ref{special:early_wosign_startcom_xs_mainpaper}, StartCom started cross-signing WoSign certificates years before being acquired by them and cross-signing continued after the acquisition.
Specifically, \emph{StartCom Certificate Authority} cross-signed the \emph{Certification Authority of WoSign} in 2006, but revoked the cross-signs in March 2017.
However, a further cross-sign of \emph{WoSign CA Limited} was not revoked before 2018.
StartCom also issued the intermediate \emph{WoSign eCommerce Services Limited}, which was cross-signed by Certplus later on, in 2011 (cf. Section~\ref{special:wosign_crosssigned_by_certplus}).
These cross-signs do not yield undesired trust paths since the StartCom roots were constrained by Mozilla's and Google's \emph{not before} constraints.
However, they show a close business relationship between StartCom and WoSign years before the (not timely reported) acquisition of StartCom by WoSign in 2015.

This behavior continued after the acquisition: \emph{StartCom Certification Authority} issued and \emph{Certification Authority of WoSign G2} cross-signed the new intermediate \emph{StartCom Class 3 OV Server CA}.

\subsection{Bootstrapping: Further Examples} %
\label{app:cross_signs:bootstrapping_further_examples}

Cross-signing enables CAs to bootstrap trust in certificates while waiting for their inclusion in root stores.
We discussed several cases in Section~\ref{sec:crosssigns_in_the_wild:bootstrapping_help}.
In the following, we provide the full list of CAs for which we identified bootstrapping cases.
In our dataset, we found bootstrapping cases for
Comodo,
USERTrust,
DigiCert,
VeriSign,
Verizon,
Cybertrust,
Let's Encrypt,
IdenTrust,
WoSign,
AffirmTrust,
GlobalSign,
Actalis,
Starfield Services,
Amazon,
T-Systems,
WellsSecure,
Dell Inc.,
SECOM,
TeliaSonera,
Unizeto,
CertiPath,
Certipost,
SCEE,
Chunghwa Telecom,
Carillon Information Security Inc.,
ORC PKI,
TAIWAN-CA.COM Inc.,
Hongkong Post
ARGE DATEN - Austrian Society for Data Protection,
CNNIC,
the U.S. Government,
the Swiss Government,
and
the Belgium Root.

\subsection{Expanding Trust and Alternative Paths: Further Examples and Comodo Details} %
\label{app:crosssigns_in_the_wild:trust_coverage_and_fallback_trust}

As noted in Section~\ref{sec:crosssigns_in_the_wild:trust_coverage_and_fallback_trust}, some root certificates are included in only some root stores.
Consequently, a single issuer may provide only limited trust.
Cross-signs by selected CA certificates can close these holes by providing trust paths to uncovered root stores.

\subsubsection{Increasing Trust Store Coverage} %
\label{app:crosssigns_in_the_wild:trust_coverage_and_fallback_trust:trust_coverage_increase_general}

We find three \xscs{} that use cross-signing to increase their root store coverage.
The intermediate \xscs{} \emph{Servision Inc.}, \emph{XiPS} and \emph{KAGOYA JAPAN Inc.} were issued by GoDaddy's \emph{ValiCert Class 1 Policy Validation Authority} in 2007, 2008, and 2009, respectively.
The ValiCert certificate provides trust paths for Mozilla, Google, and Apple, but not for Microsoft's root store.
To expand the root store coverage, all three intermediates were cross-signed by the broadly trusted \emph{Security Communication RootCA1} of SECOM end of 2012, which validated issued certificates for the Microsoft ecosystem, too.

Furthermore, the ValiCert root was removed from the root stores of Mozilla and Google around 2014 and 2015 due to the deprecation of certificates that use RSA keys with less than 2048 bit~\cite{bugzilla_936304,bugzilla_881553}.
The intermediates, however, remained trusted due to the cross-sign by SECOM.
The same holds for Apple's root store when it removed the ValiCert root in 2018.
Thus, cross-signing can not only provide a broad trust coverage but also keeps intermediates operable in the face of trust-store removals of their issuers.

Unizeto did not only cross-sign WoSign roots (cf. Section~\ref{par:crosssigns_in_the_wild:root_cs:external_root_xscs:wosign}), but also provided \emph{SSL.com} with trust on Apple devices.
Specifically, \emph{Certum Trusted Network CA} cross-signed the roots \emph{SSL.com Root CA RSA} and \emph{SSL.com RV Root CA RSA R2}.
Both roots, while created 2016 and 2017, respectively, were not included in root stores before 2018.
The cross-signs by Certum in 2018 extended trust to Apple's root store which did not include the roots before 2019.

\subsubsection{Fall-Back Trust Paths} %
\label{app:crosssigns_in_the_wild:trust_coverage_and_fallback_trust:fallback_trust_general}
\begin{figure*}[tb]
  \centering
  \includegraphics[]{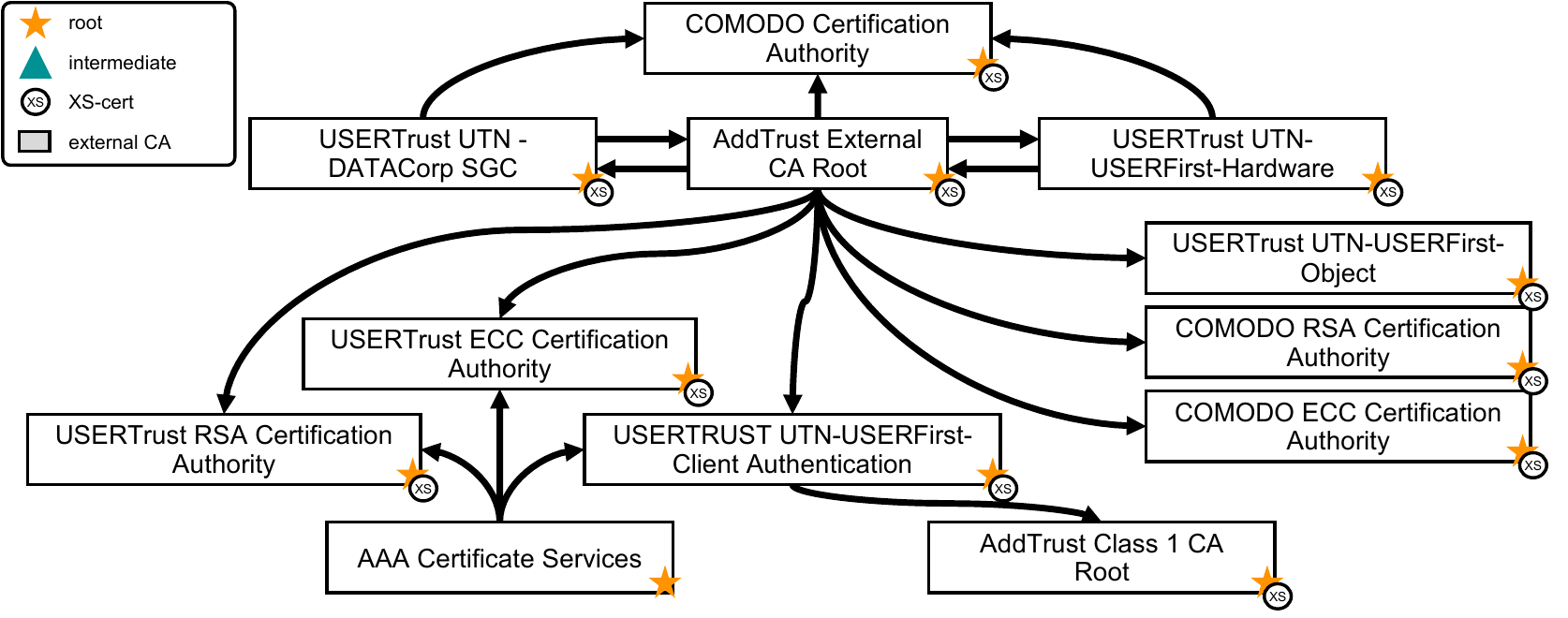}
  \caption{Internal cross-signing in the Comodo/Sectigo group.}
  \label{fig:comodo_internal}
  \Description[The figure shows cross-signs by the Comodo/Sectigo group which mainly root in AddTrust External CA Root]{
  AddTrust External CA Root, USERTrust UTN - DATACorp SGC, and USERTrust UTN- USERFirst-Hardware mutually cross-signed each other and each cross-signed COMODO Certification Authority.
  AddTrust External CA Root furthermore cross-signed USERTrust UTN-USERFirst- Object, COMODO RSA Certification Authority, COMODO ECC Certification Authority, USERTRUST UTN-USERFirst- Client Authentication, USERTrust ECC Certification Authority, and USERTrust RSA Certification Authority. The latter three were also cross-signed by AAA Certificate Services.
  USERTRUST UTN-USERFirst- Client Authentication furthermore cross-signed AddTrust Class 1 CA Root.
  All XS-certs are root XS-certs and cross-signing certificates are root XS-certs or (non-cross-signed) root certificates. All depicted XS-certs are internal XS-certs.
  }
\end{figure*}

Cross-signs that increase the root store coverage often also establish fall-back trust paths which proactively provide alternatives for unexpected revocations. 
Specifically, the SECOM cross-sign (Section~\ref{app:crosssigns_in_the_wild:trust_coverage_and_fallback_trust:trust_coverage_increase_general}) provided trust paths for certificates of Servision, XiPS, and KAGOYA JAPAN after the originally issuing ValiCert certificate was removed from Mozilla and Google root stores due to its 1024 bit RSA key~\cite{bugzilla_936304,bugzilla_881553}.

Also Entrust maintains a small internal cross-signing interconnection to establish fall-back trust paths for already broadly trusted roots: \emph{Entrust.net Secure Server Certificate Authority} cross-signed \emph{Entrust.net Certificate Authority} and \emph{Entrust Root Certificate Authority}.
The latter furthermore cross-signed \emph{Entrust Root Certificate Authority - G2} and \emph{Entrust Root Certificate Authority - EC1}.
Beyond these internal cross-signs, Entrust also received external cross-signs by today's members of the Digicert group (cf. Sections~\ref{special:entrust_xs_by_verisign}).

Apart from its cross-signing to bootstrap new roots (cf. Section~\ref{sec:crosssigns_in_the_wild:bootstrapping_help:godaddy}), GoDaddy also maintains internal \xscs{} which establish alternative trust paths for already well trusted roots.
In 2009, the bootstrapped \emph{Starfield Class 2 CA} cross-signed \emph{Starfield Services Root Certificate Authority - G2}.
\emph{Starfield Services Root Certificate Authority} added another cross-sign to this root \xsc{}.
In 2014, \emph{Starfield Class 2 CA} also cross-signed \emph{Starfield Root Certificate Authority - G2}.
Similarly in 2014, \emph{Go Daddy Class 2 Certification Authority} cross-signed \emph{Go Daddy Root Certificate Authority - G2}, establishing alternative trust paths. %
Beyond these internal cross-signs, Go Daddy's \emph{ValiCert Class 1 Policy Validation} cross-signed \emph{SECOM Security Communication Root CA1}, again establishing fall-back trust paths for an already well-trusted root.

Also GlobalSign creates fall-back trust paths via cross-signing.
Besides its incompletely revoked \xsc{} \emph{GlobalSign Extended Validation CA - SHA256 - G2} (cf. Section~\ref{special:GlobalSign_revocation_inconsistency}), GlobalSign maintains four still valid cross-signs: \emph{AlphaSSL CA - SHA256 - G2}, \emph{GlobalSign Organization Validation CA - SHA256 - G2}, \emph{GlobalSign Domain Validation CA - SHA256 - G2}, and \emph{GlobalSign CloudSSL CA - SHA256 - G3}.
The two cross-signing roots \emph{GlobalSign Root CA} and \emph{GlobalSign} provide broad trust coverage everyone for itself; hence the cross-signs establish alternative paths.

\subsubsection{Comodo's Trust-Enlarging and Fall-Back Cross-Signs} %
\label{app:crosssigns_in_the_wild:trust_coverage_and_fallback_trust:comodo}
Comodo extensively uses cross-signing to interlink root certificates within and between its subsidiary CAs as we visualize in Figure~\ref{fig:comodo_internal}.
Still, this interlinking is established by only a few certificates which cross-sign several others.
Especially the broadly trusted \emph{AddTrust External CA Root} cross-signed various root certificates of USERTrust and Comodo, establishing fall-back trust paths for them.
Almost all these cross-signs also enlarge the root store coverage, compensating for a limited root store coverage of the cross-signed root.
This effect is especially large for \emph{USERTRUST UTN-USERFirst-Client Authentication} and \emph{USERTrust UTN-USERFirst-Object} as it adds trust paths for Mozilla, Android, and the grid PKI beyond the original coverage of only iOS, OSX, and Microsoft.
Contrarily, \emph{COMODO ECC Certification Authority} and \emph{COMODO RSA Certification Authority} merely obtain new trust by the grid PKI.
Nevertheless, they still profit from the fall-back trust paths.
Before \emph{USERTrust UTN-USERFirst-Hardware} and \emph{USERTrust UTN - DATACorp SGC} ceased their operation, their mutual cross-signs with \emph{AddTrust External CA Root} also provided the latter with fall-back trust paths --- which naturally expanded to all its cross-signs.
Similarly, the broadly trusted \emph{AAA Certificate Services} root\footnote{
  The subject \emph{AAA Certificate Services} is used for a root certificate but also for an unrelated intermediate \xsc{}, i.e., the latter uses a different private key.
}
cross-signed three USERTrust roots providing each with a further fall-back trust path.

Also Comodo's intermediate \xscs{} \emph{COMODO EV SSL CA} and \emph{COMODO EV SGC CA} profit from fall-back trust paths established through cross-signing.
Comodo's very first intermediate \xsc{}, however, is \emph{LiteSSL CA} which Comodo created for \emph{Positive Software Corporation} after acquiring it in 2005.

\subsection{Switching Trust Anchor} %
\label{app:crosssigns_in_the_wild:root_switch}

\begin{figure*}[tb]
  \centering
  \includegraphics[]{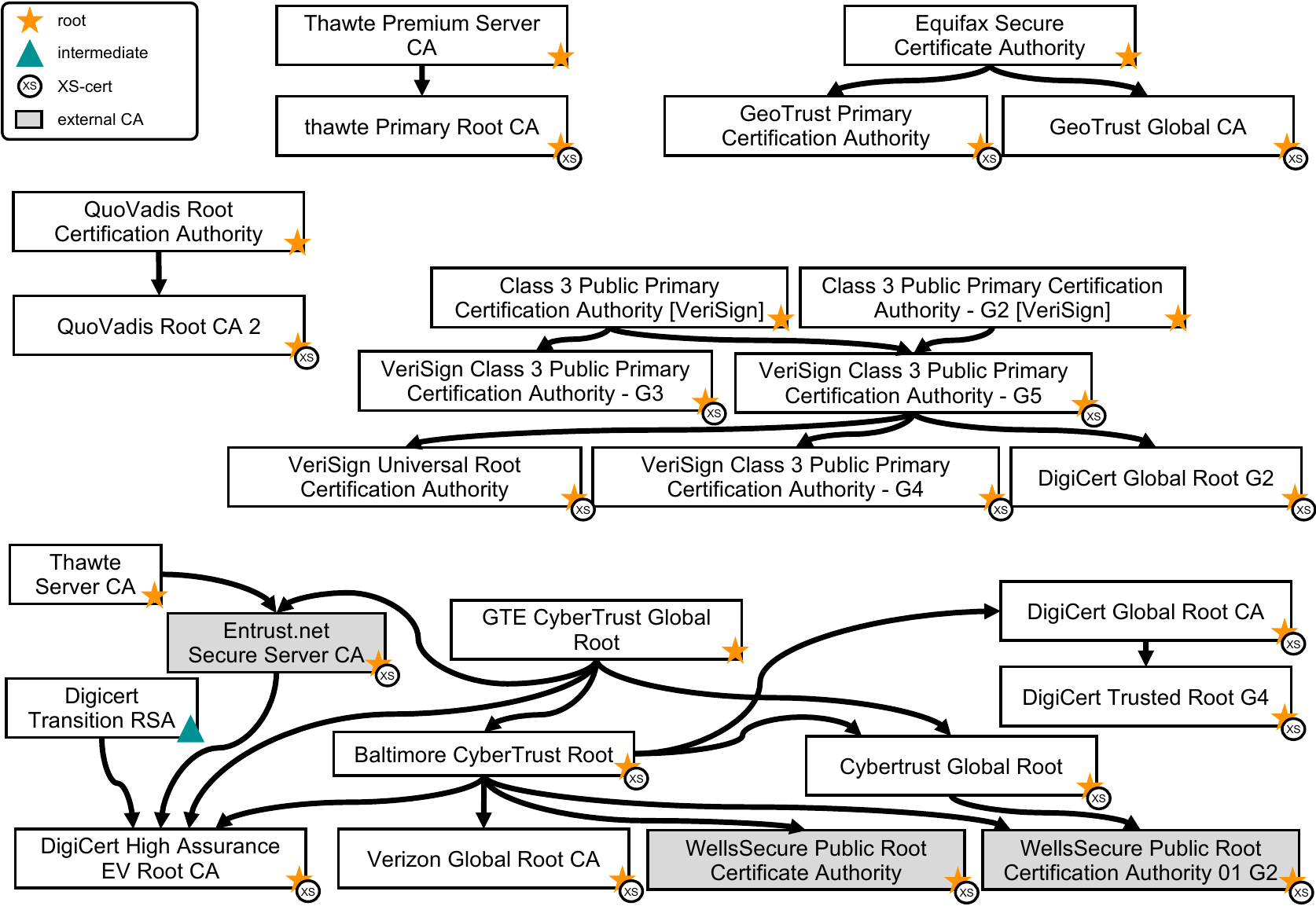}
  \caption{Root \xscs{} of Digicert and its subsidiaries: Cross-signs created prior to acquisition by Digicert form islands.}
  \label{fig:digicert_internal}
  \Description[The figure shows cross-signing island in the Digicert group. Specifically there are islands for QuoVadis, Thawte, Equifax/GeoTrust, VeriSign. There is also a larger, loosely coupled cluster for certificates from Digicert, Thawte, Cybertrust, Verizon, which also includes external certificates of Entrust and WellsSecure (by Wells Fargo)]{
  QuoVadis Root Certification Authority cross-signed QuoVadis Root CA 2.
  Thawte Premium Server CA cross-signed thawte Primary Root CA.
  Equifax Secure Certificate Authority cross-signed GeoTrust Primary Certification Authority and GeoTrust Global CA.
  Class 3 Public Primary Certification Authority [VeriSign] cross-signed VeriSign Class 3 Public Primary Certification Authority - G3 and VeriSign Class 3 Public Primary Certification Authority - G5. The latter was also cross-signed by Class 3 Public Primary Certification Authority - G2 [VeriSign] and itself cross-signed VeriSign Universal Root Certification Authority, VeriSign Class 3 Public Primary Certification Authority - G4, and DigiCert Global Root G2.
  GTE CyberTrust Global Root cross-signed the external Entrust.net Secure Server CA (also cross-signed by Thawte Server CA), DigiCert High Assurance EV Root CA (also cross-signed by Entrust.net Secure Server CA, Digicert Transition RSA and Baltimore CyberTrust Root), Baltimore CyberTrust Root, and Cybertrust Global Root.
  Baltimore CyberTrust Root also cross-signed Verizon Global Root CA, WellsSecure Public Root Certificate Authority (external), and WellsSecure Public Root Certification Authority 01 G2 (external). The latter was also cross-signed by Cybertrust Global Root.
  Except for the intermediate Digicert Transition RSA, all mentioned certificates are root XS-certs or normal root certificates.
  }
\end{figure*}

The expiry of or changing trust in certificates requires CAs to establish new trust paths to keep affected but still trustworthy certificates valid.
For example, if the issuing certificate will soon expire, the owner of an intermediate may request a different CA to issue a longer lasting intermediate with the same subject and public key.
Remember that we require certificates of a cross-sign to overlap by at least \num{121} days (cf. Section~\ref{special:xs_validity_overlap_min}) and otherwise consider it as re-issuance.
However, some trust changes happen years before expiry of the original certificates:

Symantec's \emph{GeoTrust Global CA} issued an intermediate for \emph{UniCredit} in 2012 and renewed it in 2015.
In October 2016, Actalis cross-signed this certificate.
However, the older GeoTrust issued certs were revoked the day after this issuance, rendering this to a switch to a new root rather than a typical cross-sign.

Similarly, The Spanish \emph{Autoridad de Certificacion Firmaprofesional CIF A62634068} moved from a soon expiring to a more recent root.

Also Sertifitseerimiskeskus (SK), which is the partner of Estonia for id products, used cross-signing to switch its \emph{AS Sertifitseerimiskeskus} root (expiring in 2016) to the new \emph{EE Certification Centre Root CA}.
Interestingly, the cross-sign does include a SHA3 intermediate since 2015 but no cross-sign with SHA2, limiting clients without SHA3 capabilities to the use of a SHA1 trust path element.

\subsection{Ownership Changes: Further Examples and Digicert Details} %
\label{app:cross_signs:further_examples_with_ownership_change}

In Section~\ref{sec:crosssigns_in_the_wild:ownership_changes}, we showed that cross-signs frequently outlive ownership changes which may lead missing awareness of the cross-signs and overseen problems.
We provide further examples and a detailed view on the Digicert cases in the following.

\subsubsection{Entrust Datacard} %
\label{app:cross_signs:further_examples_with_ownership_change:entrust_datacard}

Entrust Datacard is another example with cross-signs that span an ownership change: In 2016, Entrust Datacard obtained control of \emph{Affirmtrust Networking} and \emph{Affirmtrust Commercial} which were cross-signed by \emph{SwissSign Gold CA - G2} in 2009.
The cross-signs for \emph{Affirmtrust Networking} were revoked more than a year later due to a double-use of the serial number \cite{bugzilla_1404403}.
In contrast, the \emph{Affirmtrust Commercial} cross-sign was revoked only several months before it's expiry in 2019.
The corresponding root certificates, however, remain trusted.

\subsubsection{Digicert Details} %
\label{app:cross_signs:further_examples_with_ownership_change:digicert_details}
\label{app:cross_signs:digicert_details}

In Section~\ref{sec:crosssigns_in_the_wild:ownership_changes:digicert}, we outlined that Digicert inherited most of its cross-signs when acquiring CAs.
We provide the details on these cross-signs in the following.

In the Digicert group, most \xscs{} were created before the corresponding CAs were acquired by Digicert.
Thus, we predominantly find cross-signs between certificates of the same subsidiary and only very few cross-signs across Digicert subsidiaries as we illustrate in Figure~\ref{fig:digicert_internal}.
Most cross-signs originate from times of VeriSign, Verizon, and QuoVadis with Digicert only adding cross-signs among its own \emph{Digicert} roots.

VeriSign cross-signed four of its own roots (\emph{VeriSign Class 3 Public Primary Certification Authority - G3}, \emph{G4}, \emph{G5}, and \emph{VeriSign Universal Root Certification Authority}),
but also accounts for the cross-sign of \emph{thawte Primary Root CA}.
Similarly, \emph{GeoTrust Primary Certification Authority}, which started out of Equifax' security business, was cross-signed during the VeriSign ownership. Only \emph{GeoTrust Global CA} was already cross-signed when still owned by Equifax.
After acquiring VeriSign as part of Symantec in 2017, Digicert only used its control over VeriSign roots to cross-sign \emph{DigiCert Global Root G2} with \emph{VeriSign Class 3 Public Primary Certification Authority - G5}.
Otherwise, it only retained the existing cross-signs.
\label{special:entrust_xs_by_verisign}
We also find an external cross-sign that was created by VeriSign, but it did not last until the owner change to Symantec:
When Thawte was owned by VeriSign, \emph{Thawte Server} cross-signed \emph{Entrust.net Secure Server Certificate Authority}.
After expiry of these intermediates in 2003, Thawte did not renew the cross-sign, however, Entrust obtained a cross-sign by the Digicert-controlled \emph{GTE CyberTrust Global Root} in 2004.

Also most intermediate \xscs{} in the Digicert group were created by VeriSign or Symantec before Digicert obtained control over these CAs.
Specifically, VeriSign created four intermediate \xscs{} for Thawte
as well as three for itself.
After acquiring VeriSign, Symantec, which did not cross-sign any of its roots, further created \num{3} intermediate \xscs{} for its own CA name.
Only one further Symantec intermediate \xsc{} was created by Digicert after its acquisition.
Only two (three) months after the cross-signing (acquisition), Digicert revoked all certificates of this latter \xsc{}, but otherwise kept the old intermediate \xscs{} active.

A cross-sign of QuoVadis which already existed when it was acquired by WiSeKey in 2017 survived as an isolated island within the Digicert group.

Beyond these scope-wise limited cross-signing, Verizon had numerous internal and external cross-signs which became part of the Digicert group when it acquired the Verizon and CyberTrust roots from Verizon.
Internally, Verizon established multiple cross-signs across CyberTrust roots, and a Verizon root cross-sign (cf. Figure~\ref{fig:digicert_internal}).
Digicert only used \emph{Baltimore CyberTrust Root} to cross-sign \emph{DigiCert Global Root CA} and \emph{DigiCert High Assurance EV Root CA}.
The latter was additionally cross-signed by \emph{GTE CyberTrust Global Root} and \emph{DigiCert Transition RSA Root} (and \emph{Entrust.net Secure Server Certification Authority}; cf. Section~\ref{special:digicert_xs_by_entrustDatacard}).
However, more prevalent are the cross-signs of external organizations. %
First, Verizon started cross-signing root certificates of WellsFargo since 2013.
Specifically the \emph{Baltimore CyberTrust Root} cross-signed WellsFargo's \emph{WellsSecure Public Root Certificate Authority} and \emph{WellsSecure Public Root Certification Authority 01 G2} in 2013 and 2015, respectively.
The latter was already cross-signed by \emph{Verizon Global Root CA} in 2013.
All corresponding intermediates were revoked by the CA's CRL and Mozilla's OneCRL in 2017, when the roots were removed from almost all root stores (after request by WellsFargo \cite{bugzilla_1332059}), except for Apple's store which still includes \emph{WellsSecure Public Root Certificate Authority}.
Similarly, the Verizon-controlled \emph{GTE CyberTrust Global Root} cross-signed \emph{Certipost E-Trust Primary Normalized CA} providing this former only by Microsoft trusted root a broad trust coverage and creating another cross-sign that survived the acquisition by Digicert.

Verizon also cross-signed several state-controlled CAs which is potentially problematic (cf. Section~\ref{sec:crosssigns_in_the_wild:pki_barrier_breaches}).
In 2010 and 2013, \emph{Baltimore CyberTrust Root} cross-signed the \emph{Swiss Government} root which increased the formerly limited trust (Apple and Microsoft only). %
One of the cross-signs expired in 2014, the other stayed active till after the Digicert acquisition.
In 2013, \emph{Belgium Root CA2} replaced a cross-sign by (non-Digicert) \emph{GlobalSign Root} from 2007 with a further Verizon cross-sign by \emph{CyberTrust Global Root}.
This cross-sign also stayed active after the Digicert acquisition until October 2017 when it was revoked in OneCRL~\cite{bugzilla_1407559}.
Verizon also used \emph{GTE CyberTrust Global Root} and \emph{Baltimore CyberTrust Root} to cross-sign Portugal's \emph{SCEE ECRaizEstado} (formerly trusted by Microsoft only). %
This time, Digicert actively repeated the latter cross-sign shortly after obtaining control over the CyberTrust roots.
However, when the cross-sign by \emph{GTE CyberTrust Global Root} expired in August 2018, the other cross-signs were revoked due to a series of misissuances \cite{bugzilla_1432608}.
Overall, several independent CAs thus provided state-controlled CAs with large trust coverage, larger than provided by the original root certificates.

Finally, we find only a single cross-sign between Digicert's very own roots, i.e., \emph{DigiCert Global Root CA} cross-signed \emph{DigiCert Trusted Root G4}.

\subsection{Entrust: Further Examples} %
\label{app:cross_signs:entrust}

Beyond Entrusts involvement in Digicert cross-signs (cf. Section \ref{special:entrust_xs_by_verisign}), \emph{Entrust.net Secure Server CA} also cross-signed Trustwave's \emph{Secure Trust CA} and the \emph{TDC Internet Root} of the danish ISP TDC in 2006.

Furthermore, \emph{Entrust Root Certification Authority} and \emph{Entrust Root Certification Authority - G2} created the intermediate \xsc{} \emph{Entrust Certification Authority - L1M} in 2014.
The G2 root revoked the cross-sign after a month, but replaced it with a new cross-sign.

Also the intermediate \xsc{} \emph{Entrust Certification Authority - L1B} was cross-signed by \emph{Entrust.net Secure Server Certification Authority}.
When the cross-sign expired, the originally issuing \emph{Entrust.net Certification Authority (2048)} revoked the original certificate declaring it superseded.

Moreover, Entrust cross-signed its \emph{Entrust.net Secure Server Certification Authority} and \emph{Entrust Certification Authority - L1K}.
These cross-signs only yield alternative paths.

\subsection{Cross-Signing in the Grid PKI} %
\label{app:cross_signs:grid_pki}

In the grid-PKI, we do not find a real \xsc{}, but an interesting re-issuance.
Originally, \emph{ESnet Root CA 1} issued an intermediate named \emph{NERSC Online CA}.
A month before its expiry, NERSC created a corresponding root certificate instead of requesting a new intermediate.
One could argue that ESnet helped NERSC to bootstrap the trust in its later root.

\iflongversion

\subsection{Miscellaneous \xscs{}} %
\label{app:crosssigns_in_the_world:misc:intermediate}

Finally, we list less prominent cross-signs.

\subsubsection{Miscellaneous Internal Root \xscs{}} %
\label{app:crosssigns_in_the_world:misc:root:internal}

We find some more cross-signs of root certificates issued within a CA.
Slovenia used it's \emph{SI-TRUST Root} to cross-sign its \emph{sigov-ca} and \emph{sigend-ca}.
However, the corresponding root certificates are included only in Microsoft's root store.

Several CAs furthermore used an older certificate to sign new versions:
GlobalSign used its older \emph{GlobalSign Root CA} to cross-sign the new \emph{GlobalSign Root CA - R2} as well as \emph{GlobalSign Root CA - R3}.
Similarly, \emph{Sonera Class2 CA} cross-signed \emph{TeliaSonera Root CA v1} and
Chunghwa Telecom's \emph{ePKI Root Certification Authority - G2} cross-signed the \emph{ePKI Root Certification Authority}.
Notably \emph{ePKI Root Certification Authority - G2} is itself cross-signed and the root is only included in recent Microsoft stores whereas the cross-signed version is widely trusted.
Also SECOM's \emph{Security Communication EV RootCA1} and \emph{Security Communication RootCA2} were both cross-signed by \emph{Security Communication RootCA1}.
Furthermore, \emph{Certum CA} cross-signed the newer \emph{Certum Trusted Network CA},
and the \emph{DST Root CA X3}, which also cross-signed the ISRG Roots used by Let's Encrypt, cross-signed \emph{IdenTrust Commercial Root CA 1}.
Likewise, Taiwan CA cross-signed its \emph{TWCA Global Root CA} with the \emph{TWCA Root Certification Authority}
and the \emph{Hellenic Academic and Research Institutions RootCA 2015} got cross-signed by \emph{Hellenic Academic and Research Institutions RootCA 2011}.
Also the Deutsche Telekom used its established \emph{Deutsche Telekom Root CA 2} to cross-sign the \emph{T-TeleSec GlobalRoot Class 2} and \emph{T-TeleSec GlobalRoot Class 3} roots.
Finally, the \emph{TC TrustCenter Class 2 CA}, which only got into Mozilla and Android root stores for a short time, was cross-signed by \emph{TC TrustCenter Class 2 CA II} which reached a better root store coverage, although the trust does not hold up anymore.
The exactly same applies for \emph{TC TrustCenter Class 3 CA} and it's cross-signing of \emph{TC TrustCenter Class 3 CA II}.

\subsubsection{Miscellaneous Internal Intermediate \xscs{}} %
\label{app:crosssigns_in_the_world:misc:intermediate:internal}

Keynectis used its Certplus (\emph{Class 2 Primary CA}) to issue the intermediate \emph{KEYNECTIS ICS ADVANCED Class 3 CA} and cross-signed it with \emph{OpenTrust CA for AATL G1}, likewise owned by a Keynectics subsidiary.

Chunghwa Telecom created the intermediate \xsc{} \emph{Public Certification Authority - G2} using its roots \emph{ePKI Root Certification Authority} and \emph{ePKI Root Certification Authority - G2}.

Dhimyotis created the intermediate \xscs{} \emph{Certigna Services CA} and \emph{Certigna Wild CA} which provide trust paths to its \emph{Certigna} root as well as the \emph{Certigna Root CA} which is only in the root stores of Mozilla and Microsoft yet.

\subsubsection{Miscellaneous External Intermediate \xscs{}} %
\label{app:crosssigns_in_the_world:misc:intermediate:external}

Other less prominent intermediate \xscs{} contain cross-signs issued across CAs.
The \emph{Vodafone} root issued the \emph{Vodafone (Secure Sites)} intermediate in 2006.
\emph{GTE CyberTrust Global Root} cross-signed this intermediate for 2008 to 2015, providing trust paths after the expiry of the Vodafone root in 2011.
Another cross-sign by \emph{Baltimore CyberTrust Root} extended this trust to 2017, when it was revoked, or even until its expiry in December 2019 for clients that miss checking the CRL.

The \emph{USERTrust Legacy Secure Server CA} was signed by \emph{Entrust.net Certification Authority (2048)} and \emph{Entrust.net Secure Server Certification Authority} in 2009.
14 months before expiry of those intermediates, \emph{AddTrust External CA Root} issued a replacement.
Similarly, the \emph{AAA Certificate Services} was signed by \emph{Entrust.net Secure Server Certification Authority} in 2006 and cross-signed by \emph{AddTrust External CA Root} in 2008.

When \emph{GTE CyberTrust Global Root} was owned by Verizon in 2007-09, it issued the intermediates \emph{Vodafone (Corporate Domain 2009)}, \emph{Vodafone (Secure Networks)}, \emph{MULTICERT - Entidade de Certificado 001}, \emph{MULTICERT-CA 02}, \emph{LuxTrust root CA}, and Munich Re's \emph{MRG Intermediate CA 01}.
In 2012 and 2013, it cross-signed these intermediates with \emph{Baltimore CyberTrust Root} which offered a validity up to 2025 instead of only 2018 as the older root.
Several of the intermediates were revoked later on:
Both Vodafone certs were revoked in March 2017.
OneCRL only listed the newer cross-signs as the older certificates were already distrusted by removal of the Baltimore CyberTrust Root from Mozilla's root store in 2015.
This was already applied similarly for the Munich Re intermediate \xsc{} \emph{MRG Intermediate CA 01}, which was revoked in 2014, when OneCRL did not exist yet.
OneCRL included the recent cross-sign in October 2016, when the issuer of the older intermediate was already excluded from Mozilla's root store.
Similarly, the revocation of \emph{MULTICERT - Entidade de Certificado 001} in 2019 and the revocation of the \emph{LuxTrust root CA} in 2017 did not affect the older certificate as it already expired.

In 2015, Entrust issued and cross-signed the intermediate \emph{UIS-IntB} using three different CA certificates, each widely trusted.

Finally, we found re-issuances performed by the same CA but a different CA certificate which thus do not fall in our definition of cross-signs:
In 2016, \emph{SHA-1 Federal Root CA G2} reissued the intermediate \emph{CertiPath Bridge CA} whose earlier version, issued by \emph{SHA-1 Federal Root CA}, expired in 2014, thus not rendering this case a cross-sign.
A similar case holds for \emph{SAS Public CA v1} which was issued by \emph{RSA Public Root CA v1} in 2007 and reissued two weeks after expiration using \emph{RSA Security 2048 V3}.

\fi %

\bibliographystyle{ACM-Reference-Format}
\bibliography{paper}

\end{document}